\documentclass[preprint]{revtex4}
\usepackage{amsmath}
\usepackage{graphicx}
\usepackage{amssymb}
\usepackage{latexsym}
\usepackage{amsthm, amscd}
\usepackage{cancel}
\usepackage{times}
\usepackage{pstricks}
\usepackage{wasysym}
\usepackage{ulem}
\usepackage[dvips]{epsfig}
\usepackage{subfigure}
\usepackage{amssymb}

\newcommand{\dslash}{\not{\hbox{\kern-2pt $\partial$}}}
\newcommand{\td}{\tilde}
\newcommand{\bq}{\begin{equation}}
\newcommand{\eq}{\end{equation}}
\newcommand{\bqa}{\begin{eqnarray} \displaystyle}
\newcommand{\eqa}{\end{eqnarray}}
\newcommand{\nn}{\nonumber \\}
\newcommand{\bw}{\begin{widetext}}
\newcommand{\ew}{\end{widetext}}

\newcommand{\tr}{\mbox{tr}}

\newcommand{\CCC}{{\{C_1,..,C_n\}}}

\newcommand{\NS}{$NS_{D-4}$}

\begin{document}


\title{Holographic Matter : Deconfined String at Criticality}

\author{Sung-Sik Lee$^{1,2}$\vspace{0.7cm}\\
{\normalsize{$^1$Department of Physics $\&$ Astronomy, McMaster University,}}\\
{\normalsize{1280 Main St. W., Hamilton ON L8S 4M1, Canada}}\vspace{0.2cm}\\
{\normalsize{$^2$Perimeter Institute for Theoretical Physics,}}\\
{\normalsize{31 Caroline St. N., Waterloo ON N2L 2Y5, Canada}}
}

\date{\today}

\begin{abstract}

We derive a holographic dual 
for a gauged matrix model in general dimensions
from a first-principle construction.
The dual theory is shown to be
a closed string field theory which includes 
a compact two-form gauge field
coupled with closed strings in one higher dimensional space.
Possible phases of the matrix model are discussed
in the holographic description.
Besides the confinement phase 
and the IR free deconfinement phase,
there can be two 
different classes of critical states.
The first class describes {\it holographic critical states} 
where strings are deconfined in the bulk.
The second class describes {\it non-holographic critical states} 
where strings are confined due to proliferation 
of topological defects for the two-form gauge field.
This implies that the critical states of the matrix model
which admit holographic descriptions 
with deconfined string in the bulk
form novel universality classes 
with non-trivial quantum orders
which make the holographic critical states 
qualitatively distinct from 
the non-holographic critical states.
The signatures of the non-trivial quantum orders
in the holographic states are discussed.
Finally, we discuss a possibility 
that open strings emerge
as fractionalized excitations of closed strings
along with an emergent one-form gauge field  in the bulk.

\end{abstract}

\maketitle

\section{Introduction}

Extracting dynamical information on strongly interacting  
critical states of matter
is in general a hard problem in theoretical physics.
Fortunately, there are classes of strongly coupled 
quantum field theories 
whose non-perturbative dynamics can be accessed
through dual descriptions 
which become weakly coupled
when the number of degrees of freedom is large.

One such dual description that has been extensively studied
in condensed matter physics is the so-called slave-particle formulation\cite{ANDERSON,BASKARAN,SL_REVIEW}.
In this theory, a gauge redundancy is introduced 
in order to take into account dynamical constraints
imposed by strong interactions.
Unphysical states introduced in the redundant description
is projected out by a dynamical gauge field.
In the large $N$ limit, where $N$ is the number of 
flavor degrees of freedom, the dynamical gauge field
becomes weakly coupled and emerges as a low energy 
collective excitation of the system.

The slave-particle theory may be viewed as a mere
change of variables which allows one to compute dynamical properties conveniently, 
which could have been computed using a different set of variables albeit more complicated. 
However, the real power of the mathematical reformulation
lies in the fact that it allows one to  
classify various novel phases of matter
beyond the symmetry breaking scheme\cite{Wen_SL}.
In particular, those phases that support
emergent gauge boson possess subtle quantum orders
that make them qualitatively 
distinct from the conventional phases.
Because of the non-trivial quantum orders,
the phases with an emergent (deconfined) gauge boson
can not be smoothly connected to the conventional phases.
Signatures of the non-trivial quantum order include
fractionalized excitations
and protected gapless excitations (or ground state degeneracy
on a space with a non-trivial topology).

The gauge-string correspondence is another type of duality\cite{MALDACENA,GUBSER,WITTEN}.
According to the duality, 
a class of $D$-dimensional quantum field theories is dual to 
a $(D+1)$-dimensional string theory.
The question we would like to address in this paper is :
{\it Do those phases that admit holographic descriptions
in one higher dimensional space possess 
non-trivial quantum orders ? }
If so, what we call {\it holographic states} 
that can be described
in one higher dimensional space 
can not be smoothly connected 
to the conventional {\it non-holographic states}.
We claim that the answer to this question is `yes'.
The signatures of the non-trivial quantum order in holographic phases
are the emergent space with an extra dimension,
deconfined strings
and the existence of an operator 
whose scaling dimension is protected
from acquiring a large quantum correction
at strong coupling in the large $N$ limit,
even though the operator is not 
protected by any microscopic symmetry of the model.

The paper is organized in the following way.
In Sec. II, we start by reviewing the slave-particle theory
with an emphasis on quantum order in fractionalized phases.
In Sec. III, we introduce a gauged matrix model 
which will be the focus of the rest of the paper.
The model is general enough to include the $U(N)$ gauge theory.
In Sec. IV, through a first-principle derivation,
we show that the matrix model
in general dimensions is holographically dual to 
a closed string field theory in one-higher dimensional spaces.
In Sec. V, it is shown that the partition function of the original matrix model
can be interpreted as a transition amplitude between 
quantum many-loop states in the holographic description.
In Sec. VI, we show that the holographic description
has a gauge redundancy, 
and strings are coupled with a compact two-form gauge field in the bulk.
Because of the compact nature of the two-form gauge field,
topological defects for the two-form gauge field are allowed.
In Sec. VII, we discuss possible states of the matrix model.
Different states are characterized 
by different dynamics of topological defects in the bulk.
If topological defects are gapped, 
strings are deconfined in the bulk, 
and the holographic state is stable.
On the other hand, 
if topological defects are condensed,
strings are confined,
 and the bulk description is not useful anymore.
Suppressed topological defect in the holographic phase
is responsible for a non-trivial quantum order
which protects the scaling dimension of the 
phase mode of Wilson loop operators 
from acquiring a large quantum correction
at strong coupling in the large $N$ limit.
We discuss the differences between 
the holographic and non-holographic states.
The holographic critical phases can be divided further into two different classes.
In the first case, there exist only closed strings in the bulk.
In the second case, there are both closed and open strings,
where open strings emerge as fractionalized collective excitations
of closed strings.
The latter state has a yet another quantum order 
which supports an emergent one-form gauge field in the bulk.
Finally, we close with speculative
discussions on a possible phase diagram,
a world sheet description of deconfined strings,
and a continuum limit.

The present construction is beyond
the level of identifying the equations of motion in the bulk
with the beta function of the boundary theory.
We construct a full quantum theory of string in the bulk
that is dual to the boundary theory.
The construction of the dual theory makes use of the fact that loop variables associated with Wilson loops become classical objects in the planar limit of matrix models\cite{POLYAKOV80,SAKITA,MM,JEVICKI81}.
The current construction of the string field theory is directly based on the earlier works\cite{SLEE10,SLEE11}.
Compared to the the previous work on the $U(N)$ gauge theory\cite{SLEE11},
the present construction has two major improvements.
First, the extra dimension generated out of the renormalization group flow is continuous,
while the earlier construction produces a discrete extra dimension.
The infinitesimally small parameter 
associated with a continuously increasing length scale
allows one to write the bulk action in a compact form 
in this formalism.
As a result, one can readily take a continuum limit 
starting from a boundary theory defined on a lattice.
Second, the earlier construction involves infinitely many loop fields in the bulk
associated with multi-trace operators, which makes the theory highly redundant.
In the present construction, the relation between single-trace operators
and multi-trace operators are explicitly implemented.
As a result,  the dual theory can be 
written only in terms of the loop fields for single-trace operators.
Because of these improvements, 
the dual theory takes a much simpler form, 
and this transparency 
allows one to uncover deeper structures in the theory.
There also exist alternative approaches to derive holographic duals for 
general quantum field theories\cite{EMIL,DAS,Gopakumar:2004qb,KOCH,POLCHINSKI09,Douglas:2010rc,SUNDRUM}.
All these constructions including the present one are based on the notion
that the extra dimension in the holographic description 
is related to the length scale in the renormalization group flow\cite{VERLINDE,LI,HEEMSKERK10,Faulkner1010,SIN2011}.

\section{Quantum order in fractionalized phase}

In this section we review some of the key features 
of the slave-particle theory\cite{ANDERSON,BASKARAN,SL_REVIEW}
 using a pedagogical model 
introduced in Ref. \cite{GAUGE}.
We consider a model defined on the four-dimensional Euclidean hypercubic lattice,
\bqa
S  & = &  -t \sum_{<i,j>} 
\sum_{a,b} 
 \cos \left( \theta^{ab}_{i} - \theta^{ab}_j \right)    \nn
&&  - K \sum_i \sum_{a, b, c} \cos \left( 
\theta^{a b}_i + \theta^{b c}_i + \theta^{c a}_i  \right).
\label{EXCITON_MODEL}
\eqa
Here $\theta_i^{ab}$'s describe phase fluctuations of
boson fields defined at site $i$.
Each boson carries one flavor index $a$ 
and one anti-flavor index $b$
with $a,b = 1,2,...,N$.
$<i,j>$ represents nearest neighbor bonds of the lattice.
We assume that the phases satisfy the constraints
$\theta^{ab} = - \theta^{ba}$\footnote{
This model including the constraints 
can arise as an effective theory 
for exciton bose condensates 
in a multi-band insulator\cite{GAUGE}.
But we treat this model as our `microscopic model' 
for the following discussion.}.
With the constraints, 
there are $N(N-1)/2$ independent boson fields per site.
The theory has $U(1)^{N-1}$ global symmetry under which the boson fields transform as
$\theta^{ab}_{i} \rightarrow \theta^{ab}_{i} + \varphi^a - \varphi^b$.

In the weak coupling limit ($K << 1$), 
the model describes weakly coupled bosons.
As the strength of the kinetic term $t$ is increased,
there is a phase transition from the disordered phase
to the bose condensed phase.
In the disordered phase, all excitations are gapped.
In the condensed phase, there are $(N-1)$ Goldstone modes.
(At the special point of $K=0$,
there are $N(N-1)/2$ Goldstone modes
due to the enhanced symmetry).

In the strong coupling limit ($K >> 1$), 
the large potential energy
imposes an additional set of dynamical constraints,
$\theta^{a b}_i + \theta^{b c}_i + \theta^{c a}_i =0$
which is solved by a decomposition,
\bqa
\theta^{ab}_i = \phi^a_i - \phi^b_i.
\label{parton}
\eqa
Here $\phi_i^a$'s are boson fields which 
parameterize the low energy manifold.
Note that these fields carry only one flavor quantum number
contrary to the original boson fields.
The new bosons are called slave-particles (or partons).
The low energy effective action for the slave-particles 
becomes
\bqa
S & = &
  -t  \sum_{<i,j>} \left[ \sum_a e^{ i ( \phi^{a}_i - \phi^{a}_j )} \right] 
\left[ \sum_b e^{ - i ( \phi^{b}_i - \phi^{b}_j )} \right]. 
\label{EA}
\eqa
Note that this theory has a $U(1)$ gauge symmetry,
\bqa
\phi^a_i \rightarrow \phi^a_i + \varphi_i.
\eqa
This is due to the $U(1)$ redundancy
introduced in the decomposition in Eq. (\ref{parton}).
Because of the gauge symmetry, the slave-particles can not hop by themselves.
However, these particles can move in space by exchanging their positions with other particles.
For example, in Eq. (\ref{EA}), the particle with flavor $a$ can hop from site $j$ to $i$
as the particle with flavor $b$ hops from $i$ to $j$.
In this sense, they can move only through the help of other slave-particles.
One can introduce a collective hopping field
$\chi_{ij} \equiv \sum_b e^{ - i ( \phi^{b}_i - \phi^{b}_j )}$
to characterize the amplitude of this mutual hopping.
If we use this collective field, Eq. (\ref{EA}) can be written as
\bqa
S & = & - t \sum_{<i,j>,a} \chi_{ij} e^{ i ( \phi^{a}_i - \phi^{a}_j )}. 
\eqa
The magnitude of the collective field characterizes the strength of hopping,
and the phase plays the role of the U(1) gauge field
to which the slave-particles are coupled electrically.
This mapping from Eq. (\ref{EA}) 
to the U(1) gauge theory 
can be made more rigorous, by using the Hubbard-Stratonovich 
transformation\cite{GAUGE}.
Although the gauge field does not have the usual Maxwell's term,
the kinetic energy is generated once high energy modes
of the boson fields are integrated out,
which renormalizes the gauge coupling from infinity
to $g^2 \sim 1/N$.
It is clear that slave-particles can propagate coherently in space
only when the hopping field is `condensed',
and provides a smooth background.
Since the hopping field is not a gauge invariant quantity,
we need to be careful when we say that the hopping field is condensed.
This notion can be sharply characterized by examining dynamics of topological defect.

Because the U(1) gauge field is compact, 
monopole is allowed as a topological defect in the theory.
The mass of monopole is $O(N)$ for a large $N$.
Whether the slave-particles arise as low energy excitations
of the theory depends on the dynamics of monopole.
One can consider the following three different phases.

\begin{enumerate}

\item Confining phase

For a small $N$ and small $t$, 
monopoles are light,
and slave-particles are heavy.
If monopoles are condensed, 
strong fluctuations of the phase mode of the hopping field
confine the slave-particles.
Only gauge neutral composite particles,
which are nothing but the original bosons in Eq. (\ref{EXCITON_MODEL}),
 appear as
low energy excitations.
In this phase, all excitations are gapped.
This phase is adiabatically connected to the
disordered phase in the weak coupling limit.

\item Higgs phase

This is the phase 
which is electromagnetically dual to the confining phase.
The slave-particles  are condensed when $t$ is large.
As a result of the condensation of charged fields,
monopoles and anti-monopoles are connected
by vortex lines which produce
a linearly increasing potential :
monopoles and anti-monopoles are confined.
One slave-particle is eaten by the massive U(1) gauge boson,
and $(N-1)$ gapless bosons are left.
These modes are the Goldstone modes.
This phase is smoothly connected to 
the bose condensed phase in the weak coupling limit.

\item Fractionalized (Coulomb) phase

For a large $N$, the mass of monopole is large.
When both the slave-particles and monopoles are gapped,
the Coulomb phase is realized.
In this phase,  slave-particles are deconfined, and arise as (gapped) excitations of the system.
They are fractionalized modes because they carry only half the flavor quantum number 
of the original bosons.
Moreover, the U(1) gauge field arises as a gapless excitation.
It is noted that the gapless excitation 
in this phase is not a Goldstone mode.
It is not protected by any microscopic symmetry.
Saying that there is a gapless gauge boson 
in a gauge theory may sound trivial.
However, we have to remember that the gauge boson
is nothing but a collective excitation of the original boson fields.
The existence of a collective excitation which remains gapless without a fine tuning 
is actually something remarkable : 
someone who does not use the language of gauge theory
would find the origin of the gapless collective excitation
 mysterious.
It turns out that the gapless mode is protected 
by a subtle order which is not 
characterized by any symmetry breaking scheme.
This order, dubbed as quantum order\cite{Wen_SL}, is associated with suppression of topological excitation, monopole in the long distance limit.
Formally, this order can be expressed
as the emergence of the Bianchi identity $d F = 0$
in the long distance limit,
where $F$ is the field strength for the emergent
gauge field.
The key features of the non-trivial quantum order is the
presence of the fractionalized excitations 
and the emergent gauge field.
Note that slave-particles are not gauge invariant objects.
However, $\phi^a$'s become `classical' in the large $N$ limit
where non-perturbative fluctuations 
of the hopping field are suppressed.
In this regard, fractionalization is associated with 
the emergence of an `internal' space.

\end{enumerate}

 \begin{center}
 \begin{table}[h]
 \caption{}
   \begin{tabular}{| c | c | c | c| c| }
     \hline
                & slave-particle & monopole & low energy excitations  \\ \hline \hline
 Confining phase  & confined & condensed & $\theta^{ab}$  \\ \hline
Coulomb phase & deconfined & gapped   & $\phi^a$, monopole, gauge boson \\    \hline
 Higgs phase & condensed & confined  & Goldstone bosons  \\    \hline
   \end{tabular}
  \end{table}
 \end{center}

Table. I summarizes the physics 
in each phase of the boson model.
Now, we switch gear to discuss about
a matrix model and its possible phases.
We will draw a close analogy 
between the quantum order present 
in the Coulomb phase of the boson model 
and a quantum order present in the holographic phase
of the matrix model.
We will see that the holographic phase has a distinct
quantum order associated with 
the emergence of  an `external' space.

\section{Matrix model}
We start with a matrix model
defined on the D-dimensional Euclidean hypercubic lattice,
\bqa
Z = \int d U ~~ e^{-S[U]}
\label{Z}
\eqa
with the action,
\bqa
S[U] = N M^2 \sum_{<i,j>} \tr (U_{ij}^\dagger U_{ij} )
+N^2 V\left[ \frac{1}{N} W_C \right].
\label{action}
\eqa
Here $i$, $j$ are site indices in the lattice 
with lattice spacing $a$, 
and $U_{ij} = U_{ji}^\dagger$ is $N \times N$ {\it complex matrix}
defined on the nearest neighbor bond $<i,j>$.
$W_C$ is Wilson line defined on the closed oriented loop $C$,
\bqa
W_C & = & \tr \left[ \prod_{<ij> \in C} U_{ij} \right],
\label{WC}
\eqa
where the product is ordered along the path.
$V[W_C/N]$ is a function of Wilson loop operators,
\bqa
V & = & - \sum_{n=1}^\infty N^{-n}  \sum_\CCC  J_\CCC \prod_{k=1}^n W_{C_k} 
\eqa
which is, in general, non-linear in the presence of multi-trace operators.
Here $J_\CCC$'s are loop dependent coupling constants.
This theory is invariant under the $U(N)$ gauge transformation :
$U_{ij} \rightarrow V_i^\dagger U_{ij} V_j$.
Eq. (\ref{Z}) may be viewed as the partition function for a
$(D-1)$-dimensional quantum matrix model in the
imaginary time formalism.

To see that this model includes the usual $U(N)$ gauge theory,
we consider the following 
quartic action in Eq. (\ref{action}) as an example,
\bqa
N^2 V & = &  \sum_{<i,j>} \left[
- N M_0^2 ~ \tr (U_{ij}^\dagger U_{ij} )
+ N v ~ \tr ( U_{ij}^\dagger U_{ij} U_{ij}^\dagger U_{ij} )
+ v^{'} ~ \left\{ \tr ( U_{ij}^\dagger  U_{ij} ) \right\}^2 \right] \nn
&& - N J ~ \sum_\Box W_\Box,
\eqa
where $\Box$ represents unit plaquettes on the lattice.
Here $M_0^2 > 0$, $v>0$, $v^{'}>0$.
We assume that $v^{'}$ is sufficiently large compared to $J$.
The relative magnitude of $M$ and $M_0$
determines the shape of the potential for the matrix field.
For small $M_0$, $U_{ij}=0$ is the minimum,
and the system is fully gapped.
For large $M_0$, the low energy manifold is spanned by the matrices
that satisfy $U_{ij} U_{ij}^\dagger = u I$
with $u \sim \frac{ M_0^2 - M^2}{2 ( v + v^{'})}$.
In this case, the low energy effective theory becomes the $U(N)$ lattice gauge theory
with the 't Hooft coupling $\lambda \sim (J u^4 )^{-1}$.
This theory can be viewed as a `linear sigma model' for the $U(N)$ gauge theory.
Presumably, the gapped phase in the small $M_0$ limit is smoothly
connected to the confinement phase of the gauge theory.
As $M_0$ is increased further, the system can go through a phase transition
to the deconfinement phase
at a critical coupling $M_0^c$,
depending on the dimension.
If the phase transition is continuous, 
we can take the continuum limit 
by taking $a \rightarrow 0$
and $M_0 \rightarrow M_0^c$ 
such that the confining scale is fixed.

\section{General Construction}

In this section, we construct 
a holographic dual for the matrix model in Eq. (\ref{action}) 
with general potential $V$ in general dimensions.
We will follow the idea introduced in Ref. \cite{SLEE10} 
where coupling constants are lifted
to dynamical fields in the bulk space 
where the extra dimension corresponds to the
length scale of the renormalization group flow.
In the presence of multi-trace operators,
this formalism becomes rather complicated\cite{SLEE11}
because one has to introduce independent fields for infinitely
many multi-trace operators that are generated along the 
renormalization group flow.
This issue is present even though multi-trace operators are not turned on initially,
because they are generated at low energy scales in any case.
To avoid this complication, here 
we express multi-trace operators in terms of single-trace one,
by introducing a complex auxiliary field $\phi_C$ for each loop $C$ 
(see Appendix A),
\bqa
Z = \int d U
d \phi_C^{(0)} d \phi_C^{(0)*}
~~ e^{ -S_1 },
\eqa
where
\bqa
S_1 & = &
N M^2  \tr (U_{ij}^\dagger U_{ij} )
+ N^2  \phi_C^{(0)} ( \phi_C^{(0)*} - W_C/N )
+ N^2 V[ \phi_C^{(0)*} ].
\label{S1}
\eqa
Here we dropped a multiplicative numerical factor in the partition function, which is not important.
It is noted that $Z$
is well defined although
$S_1$ is not bounded from below
as a function of $\phi_C^{(0)}$ and $\phi_C^{(0)*}$.
This is because $S_1$ is complex 
and contributions from large negative $S_1$ 
is canceled because of rapid oscillation in phase.
The repeated indices $ij$ and $C$ are understood to be summed over 
nearest neighbor links and closed loops, respectively.

To perform a real space renormalization group\cite{POLCHINSKI84,POLONYI,SLEE10},
an auxiliary matrix field $\td U_{ij}$ is introduced in each link,
\bqa
Z = (N^{1/2} \mu )^{N^2 {\cal N}_l } 
\int 
d \phi_C^{(0)} d \phi_C^{(0)*}
d U d \td U ~~ e^{-S_2 },
\label{Z2}
\eqa
where
\bqa
S_2 & = & 
S_{UV}[ \phi_C^{(0)*}, \phi_C^{(0)} ] - N \phi_C^{(0)} W_C \nn
 && 
+ N M^2  \tr (U_{ij}^\dagger U_{ij} )
+ N \mu^2 \tr ( \td U_{ij}^\dagger \td U_{ij} ).
\eqa
Here ${\cal N}_l$ is the number of links in the lattice,
and
\bqa
S_{UV}[ \phi_C^{(0)*}, \phi_C^{(0)} ] & = & N^2 \Bigl\{ 
\phi_C^{(0)} \phi_C^{(0)*} + V[ \phi_C^{(0)*} ] \Bigr\}
\eqa
is an action for $\phi_C^{(0)}$.
We change the variables as
\bqa
U_{ij} & = & e^{-\alpha dz} ( u_{ij} + \td u_{ij} ), \nn
\td U_{ij} & = & e^{-\alpha dz} ( A u_{ij} + B \td u_{ij}),
\eqa
where $\alpha$ is a positive constant, 
$dz$ is an infinitesimally small parameter, and
\bq
A = -\frac{M^2}{ m \mu}, ~~ B =  \frac{m}{\mu},
\eq
with
\bqa
m^{2} &=& \frac{ M^2}{ e^{2 \alpha dz} - 1}.
\eqa
In terms of the new variables, the partition function becomes
\bqa
Z = (N^{1/2} m )^{N^2 {\cal N}_l } 
\int 
d \phi_C^{(0)} d \phi_C^{(0)*}
d u d \td u ~~ e^{ -S_3 },
\eqa
where 
\bqa
S_3 &=& 
S_{UV}[ \phi_C^{(0)*}, \phi_C^{(0)} ]  
 - N \phi_C^{(0)'} W_C^{'} \nn
&& + N  \left[ M^2 \tr ( u_{ij}^\dagger u_{ij} )
+ m^2\tr ( \td u_{ij}^\dagger \td u_{ij} ) \right] .
\eqa
Here $W^{'}_C = \tr \left[ \prod_{<ij> \in C} ( u_{ij} + \td u_{ij} )\right]$,
and $\phi_C^{(0)'}  = e^{-\alpha dz L_C} \phi_C^{(0)}$,
where $L_C$ is the length of the loop $C$.
The field $\td u_{ij}$ with the large mass $m$
has taken away
a small amount of quantum fluctuations from the original 
field $U_{ij}$, which leaves an action for $u_{ij}$
with smaller couplings $\phi_C^{(0)'}$.
Therefore, we can interpret $u_{ij}$'s as low energy fields
and $\td u_{ij}$'s as high energy fields.

\begin{figure}[h!]
\centering
      \includegraphics[height=10cm,width=8cm]{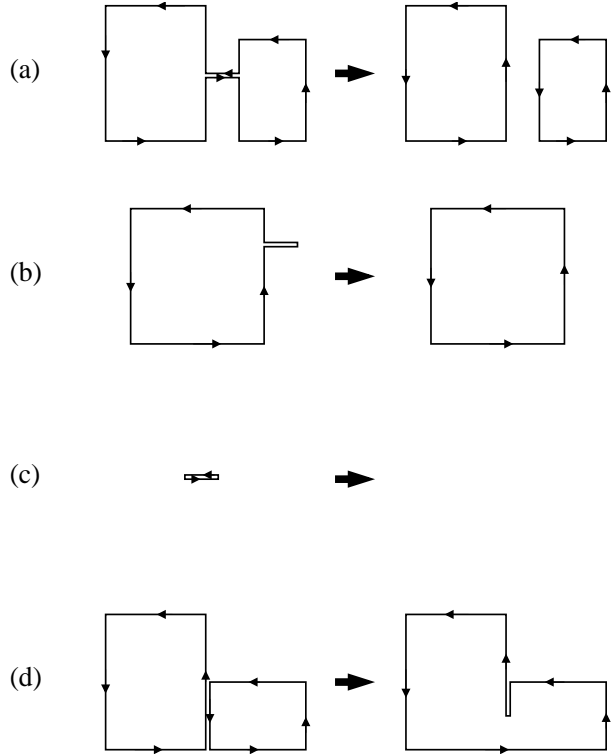}
\caption{
A loop with a self-retracting link splits into two loops (a),
becomes shorter (b), or disappears (c),
as the matrix field $\td u_{ij}$ on the link is integrated out.
In (d), two loops which share a link merge into one.
}
\label{fig:contraction}
\end{figure}

Fluctuations of $\td u_{ij}$ renormalize the 
(dynamical) couplings for the low energy field $u_{ij}$.
Integrating over $\td u_{ij}$, we obtain
\bqa
Z =\int 
d \phi_C^{(0)} d \phi_C^{(0)*}
d u  ~~ e^{ -S_4 }
\eqa
to the linear order in $dz$, where 
\bqa
S_4 &=& 
S_{UV}[ \phi_C^{(0)*}, \phi_C^{(0)} ]  
- N \phi_C^{(0)'} w_C \nn
&&
 - \frac{1}{2 m^2} F_{ij}[C_1,C_2] \phi_{[C_1+C_2]_{ij}}^{(0)'} w_{C_1} w_{C_2}
 - \frac{N}{2 m^2} G_{ij}[C_1,C_2] \phi_{C_1}^{(0)'} \phi_{C_2}^{(0)'} w_{(C_1+C_2)_{ij}} \nn
&& + N  M^2 \tr ( u_{ij}^\dagger u_{ij} )
\eqa
with $w_C = \tr \left[ \prod_{<ij> \in C}  u_{ij} \right]$.
In the third and the fourth terms, 
$ij$ runs over all nearest neighbor links,
and $C_1$, $C_2$ are understood to run over all possible loops
including null loops 
with the convention $\phi_{\emptyset_i}=0$, $\phi_{\emptyset_i}^*=1$ and $w_{\emptyset_i}=1$
for null loops,
where $\emptyset_i$ refers to the null loop at site $i$. 
Here we regard null loops at different sites as different 
loops.
By this, we can keep the combinatorics simpler.
In the third term, 
$F_{ij}[C_1,C_2]$ is a form factor that tells whether or not
two loops $C_1$ and $C_2$ are `nearest neighbors' :
$F_{ij}[C_1,C_2]=1$ if $C_1$ and $C_2$ can be merged
into one loop by adding the link $ij$ 
and rejoining the loops,
and $F_{ij}[C_1,C_2]=0$ otherwise.
$[C_1,C_2]_{ij}$ denotes the loop that is made 
of $C_1$ and $C_2$ with the addition of the link $ij$.
When both $C_1$ and $C_2$ are non-trivial loops, 
the third term describes a process 
where a loop splits into two loops (Fig. \ref{fig:contraction} (a)).
When one of the two loops is a null loop, 
it describes a process where 
a loop becomes shorter by eliminating a self-retracting link
(Fig. \ref{fig:contraction} (b)). 
When both are null loops, 
it describes a self-retracting link disappearing
(Fig. \ref{fig:contraction} (c)). 
In the fourth term, 
$G_{ij}[C_1,C_2]$ is a form factor that tells whether or not
two loops $C_1$ and $C_2$ are sharing the link $ij$ :
$G_{ij}[C_1,C_2]=1$ if $C_1$ and $C_2$ can be merged
into one loop by removing the shared link $ij$,
and $G_{ij}[C_1,C_2]=0$ otherwise.
$(C_1,C_2)_{ij}$ denotes the loop that is made 
by merging $C_1$ and $C_2$ by removing the shared link $ij$.
The fourth term describes a process where two loops 
merge into one loop
(Fig. \ref{fig:contraction} (d)). 
In the small $dz$ limit, $1/m^2 \sim O(dz)$,
and we can replace  $\phi_{C}^{(0)'}$ 
with $\phi_{C}^{(0)}$ 
in the third and fourth terms of the action
to the linear order in $dz$.

Note that double trace operators are generated for $u_{ij}$.
Another set of auxiliary fields is introduced to express
the double-trace operator in terms of single-trace operators as
\bqa
Z =\int 
d \phi_C^{(0)} d \phi_C^{(0)*}
d \phi_C^{(1)} d \phi_C^{(1)*}
d u  ~~ e^{ -S_5 },
\eqa
where 
\bqa
S_5
 &=& 
S_{UV}[ \phi_C^{(0)*}, \phi_C^{(0)} ] \nn
&& + N^2  \phi_C^{(1)} ( \phi_C^{(1)*} - w_C/N ) \nn
&&  - N^2 \phi_C^{(0)'}  \phi_C^{(1)*}  
- \frac{N^2}{2m^2} \left( F_{ij}[C_1,C_2] \phi_{[C_1+C_2]_{ij}}^{(0)}  \phi_{C_1}^{(1)*}   \phi_{C_2}^{(1)*} 
 + G_{ij}[C_1,C_2] \phi_{C_1}^{(0)} \phi_{C_2}^{(0)}  \phi_{(C_1+C_2)_{ij}}^{(1)*} \right) \nn
&& + N  M^2 \tr ( u_{ij}^\dagger u_{ij} ) \nn
 &=& 
S_{UV}[ \phi_C^{(0)*}, \phi_C^{(0)} ] \nn
&& + N^2  \phi_C^{(1)*} ( \phi_C^{(1)} - \phi_C^{(0)'}  ) \nn
&& - \frac{N^2  \alpha dz}{M^2}  
\left( F_{ij}[C_1,C_2] \phi_{[C_1+C_2]_{ij}}^{(0)}  \phi_{C_1}^{(1)*}   \phi_{C_2}^{(1)*} 
 + G_{ij}[C_1,C_2] \phi_{C_1}^{(0)} \phi_{C_2}^{(0)}  \phi_{(C_1+C_2)_{ij}}^{(1)*} \right) \nn
&& -N \phi^{(1)}_C w_C + N  M^2 \tr ( u_{ij}^\dagger u_{ij} ).
\label{S5}
\eqa
If we repeatedly apply the steps in Eqs. (\ref{Z2}) - (\ref{S5})
to the last line of Eq. (\ref{S5}) $R$ times, we obtain
\bqa
Z =\int 
\prod_{l=0}^R \left[ d \phi_C^{(l)} d \phi_C^{(l)*} \right] du
 ~~ e^{ -S_6 },
\eqa
where 
\bqa
S_6 
 &=& 
S_{UV}[ \phi_C^{(0)*}, \phi_C^{(0)} ] \nn
&& + N^2 \Bigl\{ 
\sum_{l=1}^R \Bigl[ 
\phi_C^{(l)*} ( \phi_C^{(l)} - \phi_C^{(l-1)} + \alpha L_C dz \phi_C^{(l-1)} ) \nn
&& - \frac{\alpha dz}{M^2} \left( F_{ij}[C_1,C_2] \phi_{[C_1+C_2]_{ij}}^{(l-1)}  \phi_{C_1}^{(l)*}   \phi_{C_2}^{(l)*} 
 + G_{ij}[C_1,C_2] \phi_{C_1}^{(l-1)} \phi_{C_2}^{(l-1)}  \phi_{(C_1+C_2)_{ij}}^{(l)*} \right) \Bigr]
\Bigr\} \nn
&& -N \phi^{(R)}_C w_C + N  M^2 \tr ( u_{ij}^\dagger u_{ij} ).
\label{S6}
 \eqa

What is the physical meaning of the auxiliary fields ?
In the last line of  Eq. (\ref{S6}),
we note that $\phi_C^{(R)}$ acts as a source for the 
low energy matrix field at scale $e^{-R dz}$.
The key difference from the standard renormalization group procedure
is that the source fields are dynamical fields rather than fixed constants at each scale\cite{SLEE10}.
On the other hand, the equation of motion for $\phi_C^{(R)}$
implies that 
\bqa
< \phi_C^{(R)*} > = \frac{1}{N}< w_C >. 
\eqa
Therefore, the conjugate field $\phi_C^{(R)*} $
describes the Wilson loop operator.
As we will see below, 
$\phi_C$ and $\phi_C^*$ are conjugate fields
which satisfy a non-trivial commutation relation :
sources and operators are conjugate to each other.

\begin{figure}[h!]
\centering
      \includegraphics[height=4cm,width=12cm]{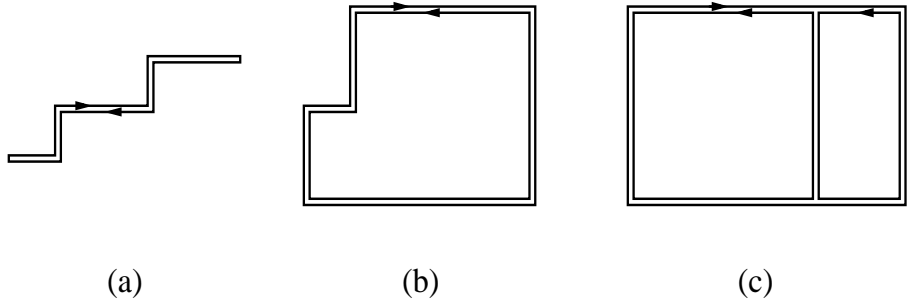}
\caption{
Examples of the diagrams that contribute to the IR effective potential to the leading order in $1/N$.
Every link should be paired with another link with 
the opposite orientation.
}
\label{fig:effectivepotential}
\end{figure}

Finally, we integrate out $u$ to obtain
\bqa
Z =\int 
\prod_{l=0}^R \left[ d \phi_C^{(l)} d \phi_C^{(l)*} \right] 
 ~~ e^{ -S_7 },
\eqa
where 
\bqa
S_7 
 &=& 
S_{UV}[ \phi_C^{(0)*}, \phi_C^{(0)} ] \nn
&& + N^2 \Bigl\{ 
\sum_{l=1}^R \Bigl[ 
\phi_C^{(l)*} ( \phi_C^{(l)} - \phi_C^{(l-1)} + \alpha L_C dz \phi_C^{(l-1)} ) \nn
&& - \frac{\alpha dz}{M^2} \left( F_{ij}[C_1,C_2] \phi_{[C_1+C_2]_{ij}}^{(l-1)}  \phi_{C_1}^{(l)*}   \phi_{C_2}^{(l)*} 
 + G_{ij}[C_1,C_2] \phi_{C_1}^{(l-1)} \phi_{C_2}^{(l-1)}  \phi_{(C_1+C_2)_{ij}}^{(l)*} \right) \Bigr]
\Bigr\} \nn
&& + S_{IR} [ \phi^{(R)}_C ].
\label{S7}
 \eqa
Here $S_{IR}$ is the effective potential given by
\bqa
S_{IR} [ \phi^{(R)}_C ] & = & -\ln \int du ~~
e^{ - N  M^2 \tr ( u_{ij}^\dagger u_{ij} ) + N \phi^{(R)}_C w_C }.
\label{S_IR}
\eqa
For a future use, we define
\bqa
V^{'}[ \phi^{(R)}_C ]  & \equiv & \frac{1}{N^2} S_{IR}[ \phi^{(R)}_C ],
\eqa
which can be computed using the strong coupling expansion,
\bqa
V^{'} [ \phi^{(R)}_C ] & = & 
- \phi^{(R)}_{C_1} M^{-L_{C_1}} {\td \delta}_{C_1,0}
- \frac{1}{2} \phi^{(R)}_{C_1} \phi^{(R)}_{C_2} M^{-\sum_{i=1}^2 L_{C_i}} {\td \delta}_{C_1+C_2,0}  \nn
&& - \frac{1}{6} \phi^{(R)}_{C_1} \phi^{(R)}_{C_2}\phi^{(R)}_{C_3} M^{-\sum_{i=1}^3 L_{C_i}} 
{\td \delta}_{C_1+C_2+C_3,0}  
- ....
\label{Vp}
\eqa
Here the delta function is defined as
\bqa
{\td \delta}_{C,0} & \equiv & \prod_{<i,j>} \delta_{ Q_{ij}[C], 0},
\eqa
where $Q_{ij}[C]$ is the U(1) charge defined on link $ij$
associated with the flux of loop $C$\cite{SLEE11}.
If the loop $C$ passes through the link $ij$ from 
$i$ to $j$ (from $j$ to $i$) $n$ times,
$Q_{ij}[C] = n (-n)$.
The first, second and third terms are 
from a self retracting loop (Fig.\ref{fig:effectivepotential} (a)),
two loops (Fig.\ref{fig:effectivepotential} (b)) 
and three loops (Fig.\ref{fig:effectivepotential} (c)), respectively.
Higher order terms can be obtained similarly.
Now we take $dz \rightarrow 0$ and $R \rightarrow \infty$ limits 
with $\beta \equiv R dz$ fixed.
Then, the partition function is written as
\bqa
Z =\int 
D \phi_C^{} D \phi_C^{*}
 ~~ e^{ -\left( S_{bulk}[\phi_C^*(z),\phi_C(z)] + S_{UV}[\phi_C^*(0),\phi_C(0)] +
 S_{IR}[\phi_C(\beta)] \right) },
\label{Z_main0}
\eqa
where 
\bqa
S_{bulk} 
 &=& 
N^2 \int_0^\beta dz \Bigl[ 
\phi_C^{*} \partial_z \phi_C^{} + \alpha L_C \phi_C^{*} \phi_C^{} \nn
&& - \frac{ \alpha }{M^2} \left( 
F_{ij}[C_1,C_2] \phi_{C_1}^{*}   \phi_{C_2}^{*} \phi_{[C_1+C_2]_{ij}}^{} 
 +  G_{ij}[C_1,C_2] \phi_{(C_1+C_2)_{ij}}^{*} \phi_{C_1}^{} \phi_{C_2}^{}  
\right) \Bigr].
\label{S_bulk0}
 \eqa
Since the partition function is independent of $\beta$,
we can take $\beta \rightarrow \infty$.
From now on, we will interpret the scale parameter $z$ as an imaginary `time'.
The dual description becomes a $(D+1)$-dimensional field theory of closed loop.
Although the action is written in terms of continuous $z$,
one should go back to the discrete version whenever
there is an ambiguity, e.g.,
when extracting boundary conditions 
by taking variations with respect to 
boundary fields.

As is the case for matrix models, there are two important parameters
that are independent with each other.
The first is $\frac{1}{N^2}$ which controls the strength of quantum fluctuations of the loop fields : the whole action including the boundary actions scales as $N^2$.
The second is the 't Hooft coupling.
In this theory, there is no unique 't Hooft coupling.
Instead there is a set of couplings defined in the space of loops, $J_\CCC$ which scales as the inverse
of the 't Hooft coupling.
Since we could have scaled out $M^2$ by redefining
$U_{ij} = U_{ij}^{'}/M$ in Eq. (\ref{action}),
the theory depends only on the 
combination $J_\CCC M^{-\sum_i L_C}$.
The small $J_\CCC$ limit is equivalent to the large $M$ limit,
which corresponds to the strong coupling limit of the matrix model where one expects to have the confinement phase.
The set of $J_\CCC$'s sets the magnitudes of loop fields in the bulk.
We will see that background loop fields, in turn, control 
the size of strings which describe small fluctuations of the loop fields.

\section{Hamiltonian picture}

\subsection{Partition function as a transition amplitude between many-body loop states}

The partition function can be viewed as
an imaginary-time transition amplitude
between many-body loop states.
To see this, we will use a rescaled loop variable
in this sub-section, 
\bqa
\Phi_C \equiv N \phi_C.
\eqa
The bulk action in the new variable becomes
\bqa
S_{bulk}
 &=& 
\int_0^\infty dz \Bigl[ 
\Phi_C^{*} \partial_z \Phi_C^{} + \alpha L_C \Phi_C^{*} \Phi_C^{} \nn
&& - \frac{ \alpha }{N M^2} \left( 
F_{ij}[C_1,C_2] \Phi_{C_1}^{*}   \Phi_{C_2}^{*} \Phi_{[C_1+C_2]_{ij}}^{} 
 + G_{ij}[C_1,C_2]  \Phi_{(C_1+C_2)_{ij}}^{*} \Phi_{C_1}^{} \Phi_{C_2}^{}  \right) \Bigr].
\label{S_bulk}
 \eqa
The action has the form 
for canonical bosonic fields, 
where $\Phi_C$ ($\Phi_C^*$)
corresponds to the coherent field associated with
the annihilation (creation) operator defined in the space of closed loops.
The annihilation and creation operators $a_C$, $a_C^\dagger$
satisfy the standard commutation relation
\bqa
\left[ a_C, a_{C^{'}}^\dagger \right] = \delta_{C,C^{'}},
\eqa
where $\delta_{C,C^{'}}$ is a Kronecker-delta function
defined in the space of loops.
Then the partition function can be written as
an imaginary-time transition amplitude,
\bqa
Z & = & \lim_{\beta \rightarrow \infty} <\Psi_f| e^{-\beta H} |\Psi_i>,
\label{transition}
\eqa 
between the initial (UV) state at $z=0$,
\bqa
|\Psi_i> & = & \int d \Phi_C^* d \Phi_C ~~
\Psi_i[\Phi_C^*, \Phi_C] |\Phi_C>,
\label{Psi_i}
\eqa
with
\bqa
\Psi_i[\Phi_C^*, \Phi_C] & = & e^{ - \Phi_C^* \Phi_C - N^2 V[ \Phi_C^*/N ]}, 
\label{wf_i}
\eqa
and the final (IR) state at $z=\infty$,
\bqa
|\Psi_f> & = & \int d \Phi_C^* d \Phi_C ~~
\Psi_f[\Phi_C^*, \Phi_C] |\Phi_C>,
\label{Psi_f}
\eqa
with 
\bqa
\Psi_f[\Phi_C^*, \Phi_C] & = & e^{ - \Phi_C^* \Phi_C - N^2 V^{'}[ \Phi_C^*/N ]}. 
\label{wf_f}
\eqa
Here 
$\Psi_i[\Phi_C^*, \Phi_C]$
and
$\Psi_f[\Phi_C^*, \Phi_C]$
are the wavefunctions of loops 
written in the coherent state basis,
\bqa
|\Phi_C> = e^{ \Phi_C a_C^\dagger } |0>,
\eqa
where $|0>$ is the vacuum in the Fock space of loops :
$a_C |0> = 0$ for all $a_C$.
(For the derivation of Eqs. (\ref{wf_i}) and (\ref{wf_f}), see Appendix. B).
The bulk Hamiltonian is given by
\bqa
H & = &
 \alpha L_C a_C^\dagger a_C 
 - \frac{ \alpha }{N M^2} \Bigl( 
F_{ij}[C_1,C_2] a_{C_1}^\dagger   a_{C_2}^\dagger a_{[C_1+C_2]_{ij}} 
 + G_{ij}[C_1,C_2] a_{(C_1+C_2)_{ij}}^\dagger  a_{C_1} a_{C_2}  
\Bigr).
\label{H}
\eqa
The first term in the Hamiltonian 
describes a tension of closed loops.
The second and the third terms are the interaction terms
which describe the processes where one loop splits into two loops,
and two loops merge into one loop, respectively,
as is shown in Fig. \ref{fig:contraction}.
We use the convention $a_{\emptyset_i} = 0$, $a^\dagger_{\emptyset_i}=1$ for null loops.
Similar loop Hamiltonians 
that describe joining and splitting processes of 
loops were considered in matrix models\cite{ISHIBASHI,HAM}.

This is an exact mapping between 
the $D$-dimensional matrix model 
($(D-1)$-dimensional quantum matrix model)
and the $(D+1)$-dimensional loop model 
(or $D$-dimensional quantum loop model).
Several remarks are in order.
First, the Hamiltonian in Eq. (\ref{H}) is a many-body Hamiltonian 
that governs the quantum dynamics of loops 
along the scale $z$ 
which is interpreted as an imaginary time.
It is noted that the Hamiltonian is not Hermitian.
Due to the cubic interaction term, 
the Hamiltonian is unbounded from below.
However, the transition amplitude in Eq. (\ref{transition}) is well defined 
because eigenvalues of the Hamiltonian are complex.
Eigenvalues with a large negative real part
in general come with a large imaginary part,
and their contributions cancel with each other 
due to oscillation in phase.
Second, the bulk Hamiltonian is {\it universal}, 
and it is independent of the details of the matrix model.
All informations pertaining to the specifics of
the matrix model 
are encoded in the initial wavefunction at $z=0$.
Third, the strength of the interaction between loops is order of $1/N$,
and loops are weakly interacting in the large $N$ limit.
Therefore, the theory becomes classical in the large $N$ limit.
Fourth, $H$ does not have any hopping term such as $a_{C_1}^\dagger a_{C_2}$ with different $C_1$ and $C_2$.
This fact will become important for gauge symmetry,
which will be discussed in Sec. V.
For earlier works on string field theories 
formulated without quadratic action,
see Ref. \cite{LYKKEN86,HOROWITZ86}.

The fact that the partition function is independent of $\beta$ has a remarkable consequence.
By taking the derivative of Eq. (\ref{transition}) 
(for a finite $\beta$) with respect to $\beta$, we obtain 
\bqa
0 & = & < \Psi_f | e^{-\beta H} H | \Psi_i >.
\eqa 
Since physical states are 
singlets of $H$,
the Hamiltonian can be viewed as a 
generator of a `gauge transformation'.
The gauge transformation corresponds to 
a reparameterization of $z$.
It is based on the fact that one can choose
different speed of renormalization group flows
at different scales without affecting the physics.
By choosing the parameter $\alpha$ to be $z$-dependent,
the reparameterization symmetry can be made explicit\cite{SLEE10}.
Here $\alpha(z)$ becomes the lapse function.
Reparameterizations of $z$ form a subgroup of
the full diffeomorphism in the $(D+1)$-dimensional space.
It would be interesting to formulate the theory
where the full diffeomorphism can be made explicit in the bulk.
Here we proceed with the present formalism 
where we choose specific time slices
along the $z$ direction.


\subsection{Wilson loop operator}

\begin{figure}[h!]
\centering
      \includegraphics[height=6cm,width=8cm]{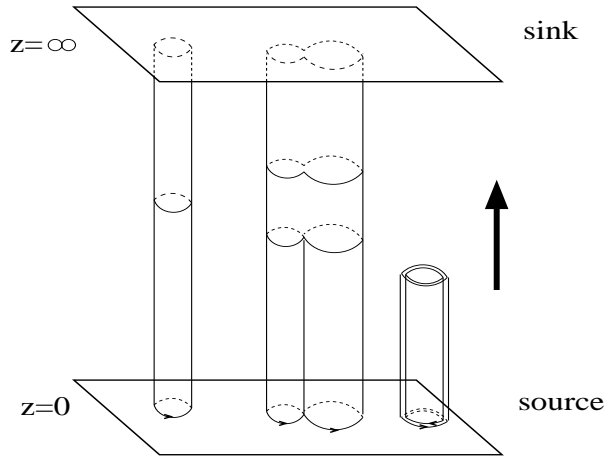}
\caption{
Loops are emitted at the UV boundary and 
propagate to the IR boundary.
Since there is no hopping term in the Hamiltonian,
loops can not move.
Instead, they can join or split following 
the processes shown in Fig. \ref{fig:contraction}.
Two loops with opposite orientations
can get pair-annihilated through 
multiple interactions.
}
\label{fig:source_sink}
\end{figure}

\begin{figure}[h!]
\centering
      \includegraphics[height=6cm,width=11cm]{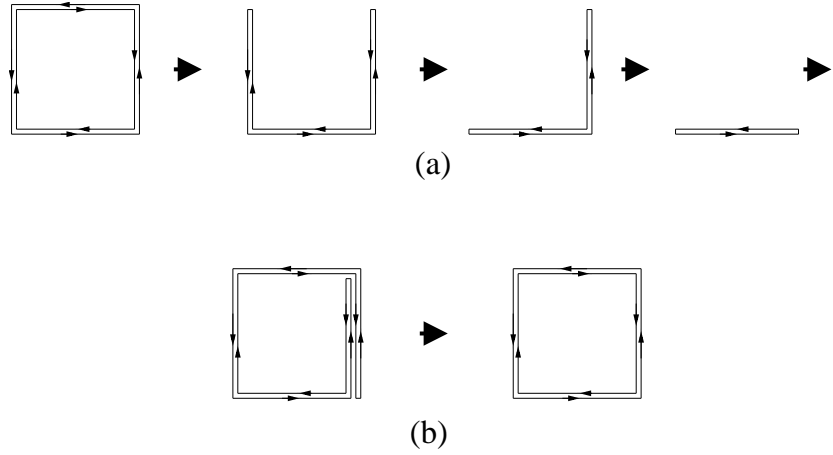}
\caption{
(a) A process where a loop and its anti-loop 
get annihilated in pair through a series of interactions.
(b) A self-retracting loop in the vacuum can split 
into a loop and an anti-loop. 
}
\label{fig:pair_annihilation}
\end{figure}

The physical picture for the transition amplitude is the following.
At the UV boundary ($z=0$), a condensate of loops 
are emitted and propagate in $z$ under the evolution governed by $H$.
The amplitude of the condensate is $<\Phi_C> \sim O(N)$.
This can be seen from the fact that the action for the unscaled loop fields
$\phi_C$ has $N^2$ as an overall prefactor, which 
implies $<\phi_C> \sim O(1)$.
Loops can join and split 
through the interactions as is illustrated in Fig. \ref{fig:source_sink}.
A loop $C$ and its anti-loop $\bar C$,
the loop with the opposite orientation, 
can get pair-annihilated through a series 
of interactions as is shown in Fig. \ref{fig:pair_annihilation} (a).
Moreover, a self-retracting loop can become
a loop and an anti-loop 
as is shown in Fig. \ref{fig:pair_annihilation} (b).
As it will be shown in Sec. VII. A, 
loop fields for self-retracting loops
have non-zero vacuum expectation values in the bulk.
Therefore, a pair of loop and anti-loop
can be created out of vacuum.
This means that two loops with the opposite orientations
act as particle and anti-particle in a relativistic
field theory.
Finally, those loops emitted at the UV boundary
are absorbed at the IR boundary.
In this sense, the UV boundary is 
a source of loops,
and the IR boundary is a sink.

\begin{figure}[h!]
\centering
      \includegraphics[height=6cm,width=8cm]{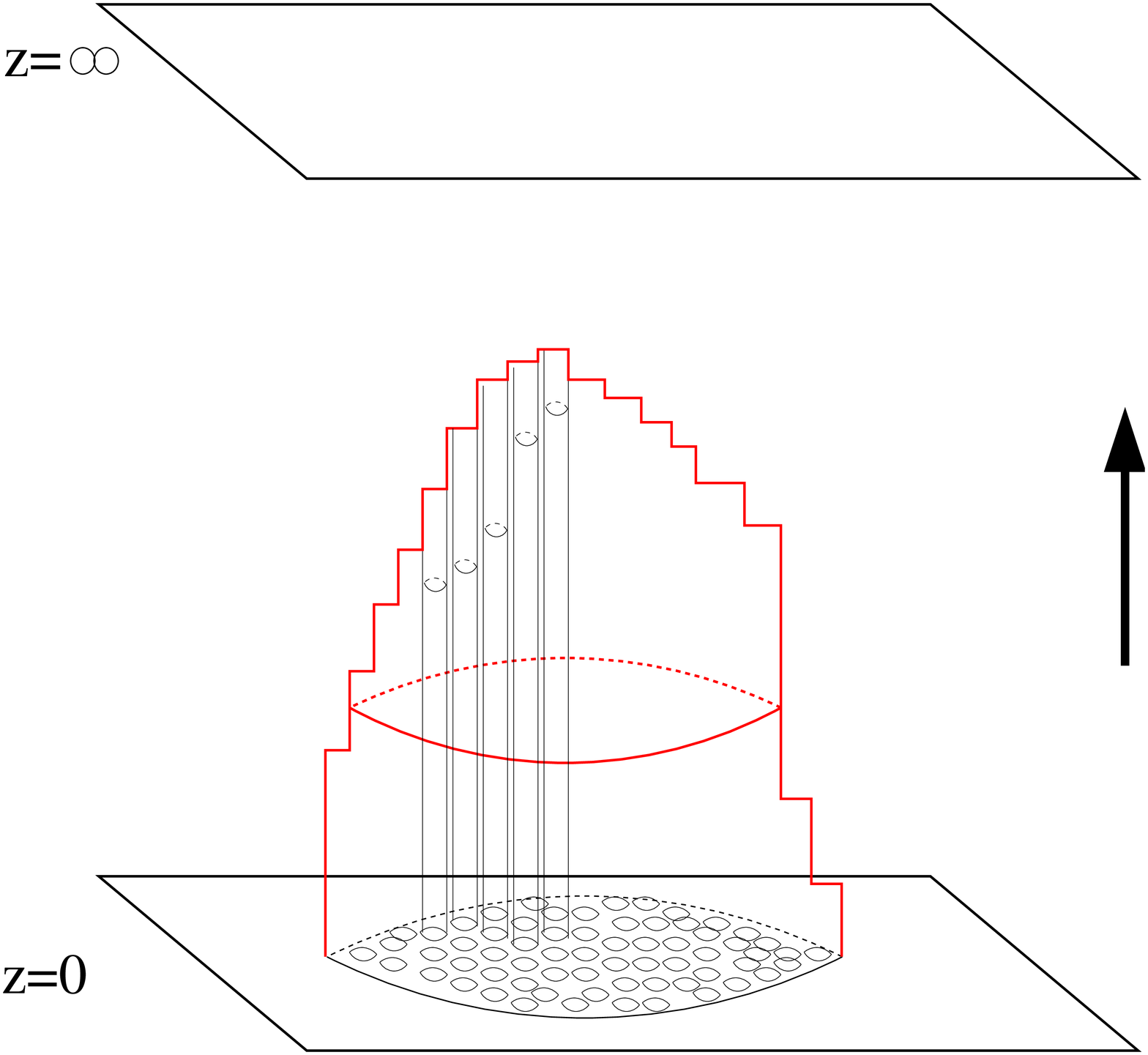}
\caption{
A world sheet formed by multiple processes where
a large loop absorbs many small loops at different stages
to change its shape to disappear before
it reach the IR boundary.
}
\label{fig:Wilson_loop}
\end{figure}

Now let us consider a Wilson loop operator for a loop $C$ which 
is much larger than the size of Wilson loops for which 
sources are turned on at the UV boundary.
The expectation value of the Wilson loop operator 
is given by the one-point function,
\bqa
<W_C> = \frac{1}{N} \left. \frac{\partial  \ln Z}{\partial J_C} \right|_{J_C=0}
 =  \lim_{\beta \rightarrow \infty}
\frac{ <\Psi_f| e^{-\beta H} a_C^\dagger |\Psi_i> }{ <\Psi_f| e^{-\beta H} |\Psi_i> }.
\eqa
If $M$ is large, loops propagate independently in the bulk.
To the zeroth order in $1/M$, the loop $C$ propagate to the sink 
along the straight path.
However, this configuration vanishes as $e^{- \alpha \beta L_C}$
in the large $\beta$ limit because of the tension.
In order for the expectation value to survive,
the large loop $C$ should absorb other smaller loops 
from the condensate to disappear 
before it reaches the IR boundary.
Then the evolution of the Wilson loop 
forms a world-sheet in the bulk.
One such configuration is shown in Fig. \ref{fig:Wilson_loop}.
Then the expectation value is given by 
the sum over all world-sheets of the Wilson loop.

Since the interaction between loops is $O(1/N)$,
loops become classical in the large $N$ limit.
This implies factorization of Wilson loop operators in the 
large $N$ limit,
\bqa
\left< \prod_{k=1}^n W_{C_k} \right> 
& = &  \prod_{k=1}^n \left< W_{C_k} \right> + O(N^{n-2}).
\eqa

\section{Gauge symmetry}

The absence of the hopping term in the Hamiltonian
has a deep origin : 
the loop field theory has a gauge symmetry.
Note that this gauge symmetry is not related to the $U(N)$
gauge symmetry of the original matrix model.
Loop fields are singlets for the $U(N)$ gauge symmetry.
In this section, we examine the consequences of the  new gauge symmetry carefully.
From now on, we return to the unscaled loop variable $\phi_C \equiv \frac{\Phi_C}{N}$.

The bulk action in Eq. (\ref{S_bulk0}) is invariant under 
the time-independent transformation generated by $Q_{ij}$ at each link
\bqa
\phi_C & \rightarrow & ~e^{i \theta_{\mu}(i) Q_{ii+\mu}[C]} ~\phi_C,
\label{loop_gauge_transformation}
\eqa
where $i$ is summed over all sites,
$\mu$ is summed over $D$ directions of nearest neighbor links, and 
$\theta_{\mu}(i)$ is a time-independent angle defined on the link $<i,i+\mu>$.
The IR boundary action respects the symmetry,
but the UV action does not.
This is because the UV potential
\bqa
V[\phi_C^*] & = & - \sum_{n=1}^\infty 
\sum_\CCC
J_{C_1,C_2,..,C_n} \left[ \prod_{k=1}^n \phi_{C_k}^* \right]
\eqa
includes sources $J_{C_1,C_2,..,C_n}$ which
explicitly break the symmetry.
It is useful to view $J_{C_1,C_2,..,C_n}$ as an expectation value
of another dynamical loop field.
Then, the full theory  is invariant if we allow the UV source to transform as
\bqa
J_{C_1,C_2,..,C_n} & \rightarrow & 
~e^{i \theta_{\mu}(i) \sum_{k=1}^n Q_{ii+\mu}[C_k]} ~J_{C_1,C_2,..,C_n}.
\eqa

This time-independent symmetry can be lifted to a full space-time gauge symmetry
by introducing temporal components of a 
two-form gauge field $B_{M N}$ in the bulk with 
$M,N = z,1,2,...,D$,
\bqa
S_{bulk} 
 &=& 
N^2 \int_0^\beta dz \Bigl[ 
\phi_C^{*} \Bigl(
\partial_z + i  Q_{ii+\mu}[C] B_{\mu z}(i,z) \Bigr) \phi_C^{} 
+ \alpha L_C \phi_C^{*} \phi_C^{} \nn
&& - \frac{ \alpha }{M^2} \left( 
F_{ij}[C_1,C_2] \phi_{C_1}^{*}   \phi_{C_2}^{*} \phi_{[C_1+C_2]_{ij}}^{} 
 +  G_{ij}[C_1,C_2] \phi_{(C_1+C_2)_{ij}}^{*} \phi_{C_1}^{} \phi_{C_2}^{}  \right) \Bigr],
\label{S_bulk_gauged}
 \eqa
where $B_{\mu z}(i)$ with $\mu=1,2,..,D$ 
are the temporal components of the two-form gauge field
defined at each spatial link.
This two-form gauge field 
is the Kalb-Ramond gauge field\cite{KR}.
Now the full theory is invariant under 
the space-time dependent gauge transformation with
\bqa
\phi_C(z) & \rightarrow & ~e^{i \theta_{\mu}(i,z) Q_{ii+\mu}[C]} ~\phi_C(z), \nn
B_{\mu z}(i,z) & \rightarrow & B_{\mu z}(i,z) + 
\Bigl[ \theta_z(i+\mu,z) - \theta_z(i,z) \Bigr] 
- ( \partial_z \theta_{\mu}(i,z)),
\label{gt}
\eqa
where $\theta_z(i,z)$ is a temporal gauge parameter defined at each site.
This is the discrete version of the usual gauge transformation for the two-form field,
$B_{MN} \rightarrow B_{MN} + \partial_M \theta_N 
- \partial_N \theta_M$.
Note that introducing the temporal components of the two-form gauge field into the theory
doesn't do anything except for making the gauge symmetry more explicit.
This can be understood from the fact that 
one can reproduce the original action in Eq. (\ref{S_bulk0}) 
by choosing the temporal gauge with $B_{\mu z}=0$.
This can be done by choosing
\bqa
\theta_\mu(i,z) = \int_0^z dz^{'} B_{\mu z}(i,z^{'})
\eqa
with $\theta_z(i,z)=0$  in Eq. (\ref{gt}).
The temporal components can be completely gauged away 
because they are pure gauge degrees of freedom 
in the presence of boundaries.
This is in contrast to the case with the periodic boundary condition, where 
the time independent component of the temporal gauge field can not be gauged away.

\begin{figure}[h!]
\centering
      \includegraphics[height=5cm,width=4.5cm]{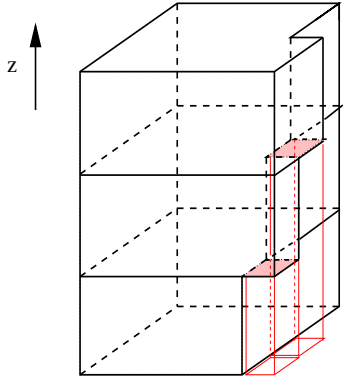}
\caption{
Loops can change their shapes and move in space like amoebas,
by absorbing or emitting small loops.
}
\label{fig:loop_propagation}
\end{figure}

As a result of the gauge symmetry,
there is no quadratic hopping term for loops in the Hamiltonian.
However, this does not necessarily mean that 
loops are always localized in space.
Loops can change their shapes and move in space
by absorbing or emitting other loops. 
For example, Fig. \ref{fig:loop_propagation} shows a loop changing its shape
by absorbing two small loops.
Therefore, loops can propagate with the help of other loops.
If loop fields are `condensed', whose precise meaning will become clear in a moment,
the condensate provides a coherent background on which
other loops can propagate.
Loops propagate `on the shoulders of other loops'
to explore the bulk space.
This is analogous to the the slave-particle theory 
discussed in Sec. II.
One difference is that 
loop fields themselves play the role of `hopping fields' for other loops,
while in slave-particle theory the hopping field 
is a bi-linear of slave-particle fields. 
The difference originates from the fact that 
loops are extended objects while
slave-particles are point objects.
Only when the condensates of loop fields are `coherent',
the bulk space is regarded as a well defined extended space by loops.
Otherwise, loops are more or less localized in space.
In this sense, an extended space emerges in the bulk
as a dynamical feature of a phase 
where loop fields form coherent condensates.

When do loop fields become coherent ?
To make this notion more precise, we first note that
the phase modes of complex loop fields $\phi_C = |\phi_C| e^{i b_C}$
play the role of the spatial components of the two-form gauge field.
To see this, suppose that the loop field has a background value $< \phi_C >$.
Then the cubic interaction generates a quadratic hopping term,
\bqa
-\frac{\alpha < \phi_C >}{M^2} a_{C+C^{'}}^\dagger a_{C^{'}}.
\eqa
The amplitude of $<\phi_C>$ is the strength of the hopping,
and the phase determines the geometric phase acquired 
when the loop $C^{'}$ hops to $C^{'}+C$.
Therefore $b_C$ plays the role of the spatial components
of the two-form gauge field to which
loops are electrically coupled.
Note that the two-form gauge field is 
also a part of dynamical loop fields.
We identify 
\bqa
b_C = \int_{A_C} B,
\eqa
where $B_{\mu \nu}$ is the spatial components of the 
two-form gauge field
and the integration is over an area $A_C$ enclosed by the loop $C$.
Let us focus on the loops with unit plaquettes
in which case we take $A_C$ as the surface spanned by the unit plaquette.

\begin{figure}[h!]
\centering
      \includegraphics[height=16cm,width=14cm]{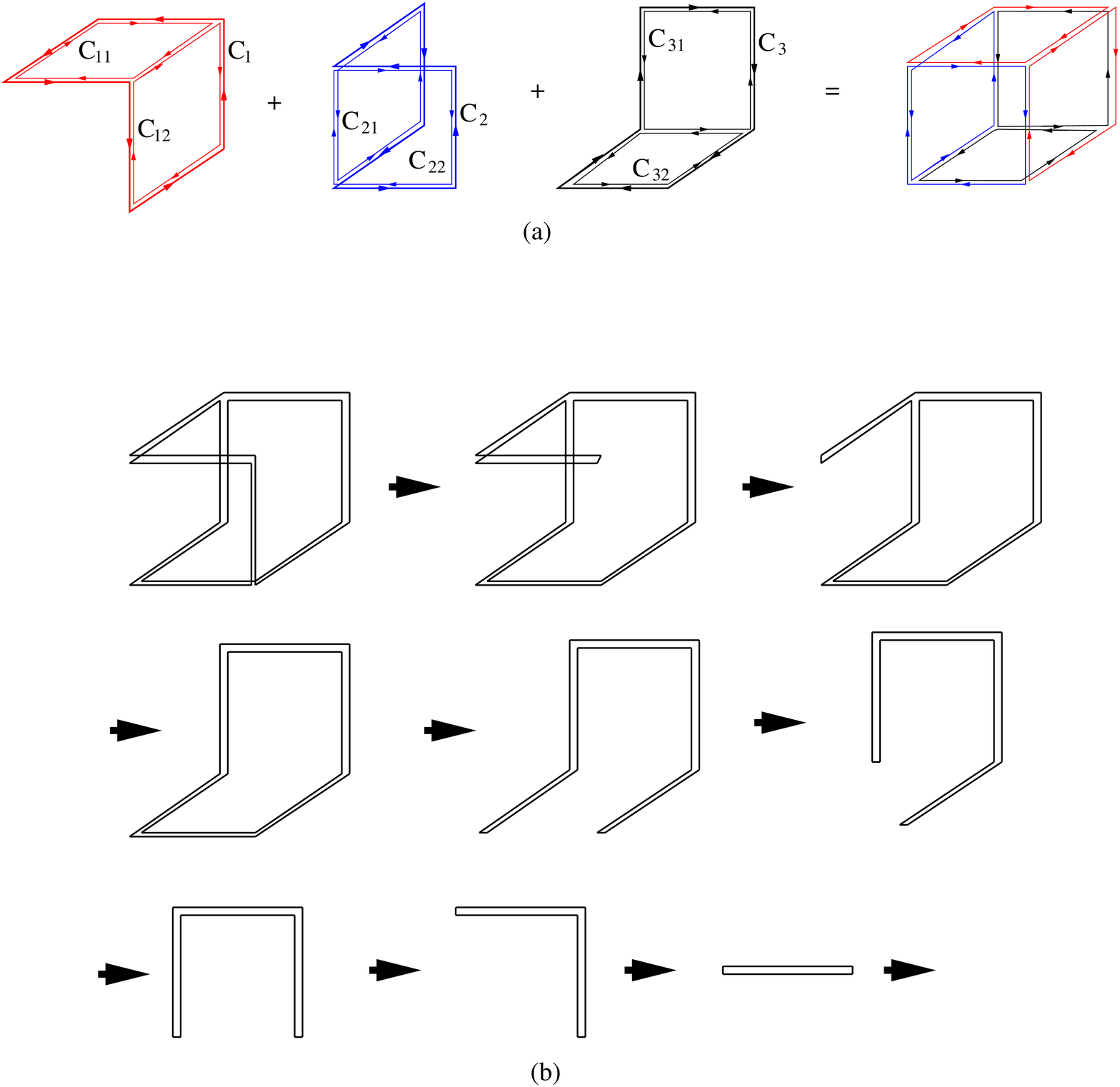}
\caption{
Kinetic energy for the two-form gauge field
generated by heavy loop fields.
(a) Three vertices each of which involves
one loop with length $6$ and two loops with length $4$
can generate a term for the six loops with length $4$, 
once the loop fields with length $6$ are integrated out.
The resulting term for the small loops
becomes the kinetic energy for the two-form gauge field.
(b) In integrating out the loop fields with length $6$ in (a),
one has to introduce a series of nine vertices.
Each step depicts a process of adding a new vertex
and integrating out one or two loop fields : 
two fields for the first and the fourth steps, 
and one loop field for the other steps.
}
\label{fig:twoform_kinetic}
\end{figure}

Although the two-form gauge field does not have the bare action,
it acquires the kinetic energy from quantum fluctuations.
This is similar to the way
that the Maxwell's term is dynamically generated
for the auxiliary gauge field 
in the slave-particle theory 
as discussed in Sec. II.
The gauge coupling for the two-form gauge field
is renormalized to $O(1/N^2)$.
This can be understood by integrating out 
`heavy' loop fields to obtain an effective 
action for `light' loop fields in the bulk.
It is easiest to see the generation of the kinetic energy
in the large $M$ limit,
where we can use $1/M$ as an expansion parameter.
The 'mass' of a loop field is proportional 
to the length of the loop because of the tension.
We integrate out loop fields with $L > 4$ 
and obtain an action for the loop fields
with $L \leq4 $.
In particular, we focus on the effective action
for the shortest non-self-retracting loops with $L_C=4$
whose phase modes can be viewed as the spatial components
of the two-form gauge field on unit plaquettes.
For simplicity, 
we choose the temporal gauge with $B_{\mu z}=0$
and the scale of $z$ to set $\alpha=1$.

Let us consider three vertices in Eq. (\ref{S_bulk_gauged}).
Each vertex has the form 
$\frac{N^2}{M} \phi_{C_i}^* \phi_{C_{i1}} \phi_{C_{i2}}$
with $i=1,2,3$,
where $C_i$'s have length $6$ and 
$C_{il}$'s with $l=1,2$
have length $4$
as is represented in Fig. \ref{fig:twoform_kinetic} (a).
They describe the processes 
where a loop on a unit plaquette with  sides $\mu, \nu$
 merges with a loop on a unit plaquette with sides $\nu, \lambda$
to form a loop with with length $6$.
Here we interpret 
$\phi_{C_i}$'s as heavy fields and 
$\phi_{C_{il}}$'s with $l=1,2$ as light fields.
In particular, the phase modes of $\phi_{C_{il}}$ 
represents the two-form gauge field defined on 
each plaquette.
Now we integrate out the heavy loop fields 
using the quadratic action. 
Because this quadratic action has the local $U(1)$ symmetry
in the loop space, $\phi_C \rightarrow e^{i \varphi_C} \phi_C$,
we need to introduce a series of vertices 
in order to saturate $\phi_C^*$ with $\phi_C$ 
and obtain a non-vanishing result.
A minimum path to saturate all heavy fields
is shown in Fig. \ref{fig:twoform_kinetic} (b).
In the first step, we add a vertex 
of the type $\frac{N^2}{M^2} \phi_{(C_1+C_2)_{ij}}^* \phi_{C_1} \phi_{C_2}$, where $(C_1+C_2)_{ij}$ is a loop
that results from merging $C_1$ and $C_2$ by removing
one shared bond.
Integrating out $\phi_{C_1}$ and $\phi_{C_2}$,
we obtain the loop fields in the second configuration
in Fig. \ref{fig:twoform_kinetic} (b).
In the second step, we use a vertex 
$\frac{N^2}{M^2} \phi_{(C_1+C_2)_{ij}^{'}}^*  \phi_{(C_1+C_2)_{ij}}$,
and integrate out $\phi_{(C_1+C_2)_{ij}}$.
In this step, the merged loop in the first step become a shorter loop 
 $(C_1+C_2)_{ij}^{'}$ by eliminating one self-retracting link.
The remaining steps can be understood in a similar way.
In total, nine vertices and eleven loop propagators are needed.
(Note that each of the first and fourth steps 
introduces two propagators because two loop fields
are integrated out in those steps, 
while all the others involve only one propagator.)
Each vertex contributes $\frac{N^2}{M^2}$ 
and each propagator contributes $\frac{1}{N^2}$.
Combined with the factor $\left( \frac{N^2}{M^2} \right)^3$ from
the original three vertices,
we obtain an action for the light loop fields,
\bqa
S_{eff} \sim - \int dz \sum_{\mbox{cubes}} 
\frac{N^2}{M^{24}}
\left[ 
\prod_{i=1}^3 \phi_{C_{i1}} \phi_{C_{i2}} 
+
\prod_{i=1}^3 \phi_{\bar C_{i1}} \phi_{\bar C_{i2}} 
\right],
\eqa
where the summation is over all cubes 
in the $D$-dimensional lattice.
The second term is from the same process 
for the anti-loops.
The loop field for the anti-loop, $\phi_{\bar C}$ 
is in priori independent of $\phi_C$.
However they are dynamically mixed.
Because of pair-annihilation 
and pair-creation processes of loops and anti-loops 
as is shown in Fig. \ref{fig:pair_annihilation},
the effective action should include terms of the form,
$\phi_C \phi_{\bar C}$ and 
$\phi_C^* \phi_{\bar C}^*$.
As a result, the phase modes of $\phi_C$
and $\phi_{\bar C}$ are locked.
Only the anti-symmetric mode
with $b_{\bar C} = - b_{C}$ remains
gapless in the presence of mixing.
If $\phi_{C_{il}}$'s have finite amplitude $\phi_0$,
this gives the standard `magnetic' term for the two-form
gauge field defined on each cube of the lattice 
\bqa
-\frac{1}{g_{KR}^2} \int dz \sum_{\mbox{cubes}}
\cos \left[ 
a^3 (
\Delta_\mu B_{\nu \lambda}
+ \Delta_\nu B_{\lambda \mu}
+ \Delta_\lambda B_{\mu \nu} )
\right],
\label{Kin_B}
\eqa
where we 
use the fact that 
$b_\Box = a^2 B_{\mu \nu}$ for 
a unit plaquette with sides $\mu$, $\nu$.
The finite derivative is defined as
$\Delta_\mu B_{\nu \lambda} 
\equiv
\frac{B_{\nu \lambda}(x+\hat x^\mu)- B_{\nu \lambda}(x)}{a}$.
Here $g_{KR}^2 \sim \left[ \frac{N^2}{M^{24}} \phi_0^6 \right]^{-1}$
is the renormalized coupling for the Kalb-Ramond (KR) two-form gauge field.

Now we turn our attention to the `electric' term
which involves the time-derivative of the gauge field.
We consider the quadratic action for 
the loop fields on unit plaquettes,
\bqa
N^2 \left[
\phi_C^* (\partial_z + L_C) \phi_C
+
\phi_{\bar C}^* (\partial_z + L_{\bar C}) \phi_{\bar C}
\right].
\eqa
If we integrate out the amplitude fluctuations 
of the loop fields, the time derivative term will be
generated for $b_C$.
Because of the dynamical constraint $b_{\bar C} = - b_C$
caused by the mixing between $\phi_C$ and $\phi_{\bar C}$,
the linear time derivative term is canceled,
leading to the second derivative term,
\bqa
N^2 a^4 \int dz \sum_\Box (\partial_z B_{\mu \nu})^2
\label{Kin_E}
\eqa
for each plaquette.
Eqs. (\ref{Kin_B}) and (\ref{Kin_E}) 
represent the full
kinetic energy term for the two-form gauge field
in the temporal gauge.

\begin{figure}[h!]
\centering
      \includegraphics[height=7cm,width=8cm]{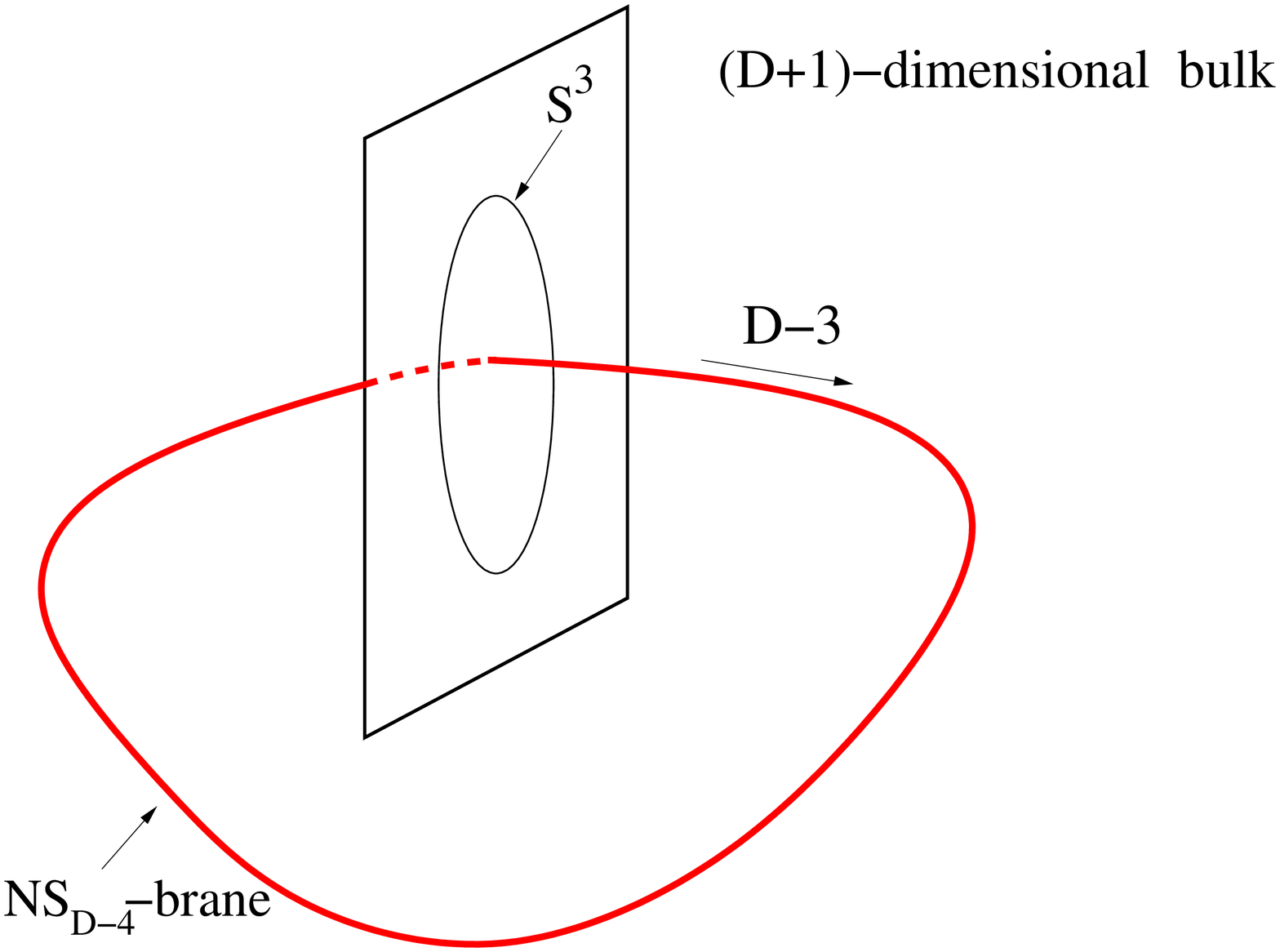}
\caption{
An \NS-brane in the $(D+1)$-dimensional space.
Around an \NS-brane, there is 
a net flux of $2\pi$ for the three-form flux :
$\int_{S^3} H = 2\pi$.
When the tension of the \NS-brane is positive,
\NS-branes generated out of quantum fluctuations
remain small.
When the tension becomes negative,
\NS-branes are proliferated in the bulk.
}
\label{fig:NS}
\end{figure}

Because of the gauge symmetry, 
the mass term is not allowed for the
two-form gauge field in the bulk.
However, the gauge symmetry
does not automatically imply 
that the two-form field arises as
a massless excitation in the bulk.
This is because of the compactness of the phase mode
: $b_C \sim b_C + 2 \pi$,
which allows for a topological defect
to exist as a magnetic excitation of the gauge field.
In the presence of topological defects,
the field strength $H = d B$ does not 
satisfy  the Bianchi identity $d H=0$.
In $(D+1)$-dimensional space-time, 
the topological defect which carries 
a magnetic charge is a $(D-4)$-brane, 
which is a $(D-3)$-dimensional object in space-time.
We call this object $NS_{D-4}$ brane\footnote{
This name has been borrowed from the NS$_5$ brane
which is the magnetically charged object
for the Kalb-Ramond two-form gauge field
in the ten dimensional superstring theory.
}.
Around the \NS-brane, there is a net $2\pi$ flux
for the three-form flux,
\bqa
d H(x) = 2 \pi \int d\xi^{D-3} \delta^{D+1}(x-\xi),
\eqa
where $\xi$ is the coordinate of the NS-brane
embedded in the $(D+1)$-dimensional space,
and $d \xi^{D-3}$ is the oriented volume element
of the brane.
This is illustrated in Fig. \ref{fig:NS}.
Note that the Dirac quantization condition 
between the charge carried by loop fields, 
which is set to be $1$ as 
can be seen from Eq. (\ref{loop_gauge_transformation}),
and the charge of the \NS-brane
is automatically satisfied.
This follows from the fact that the phase $2\pi$
on a unit plaquette is invisible to loop fields.
In $D>4$, this is a brane extended along $(D-4)$-directions in space at a given time slice with fixed $z$.
In $D=4$, this is a point-like particle.
In $D=3$, this is an instanton which is localized both in space and time.
Whether loop fields provide a coherent background for other loops 
is determined by dynamics of $NS_{D-4}$-branes.

\begin{figure}[h!]
\centering
      \includegraphics[height=8cm,width=10cm]{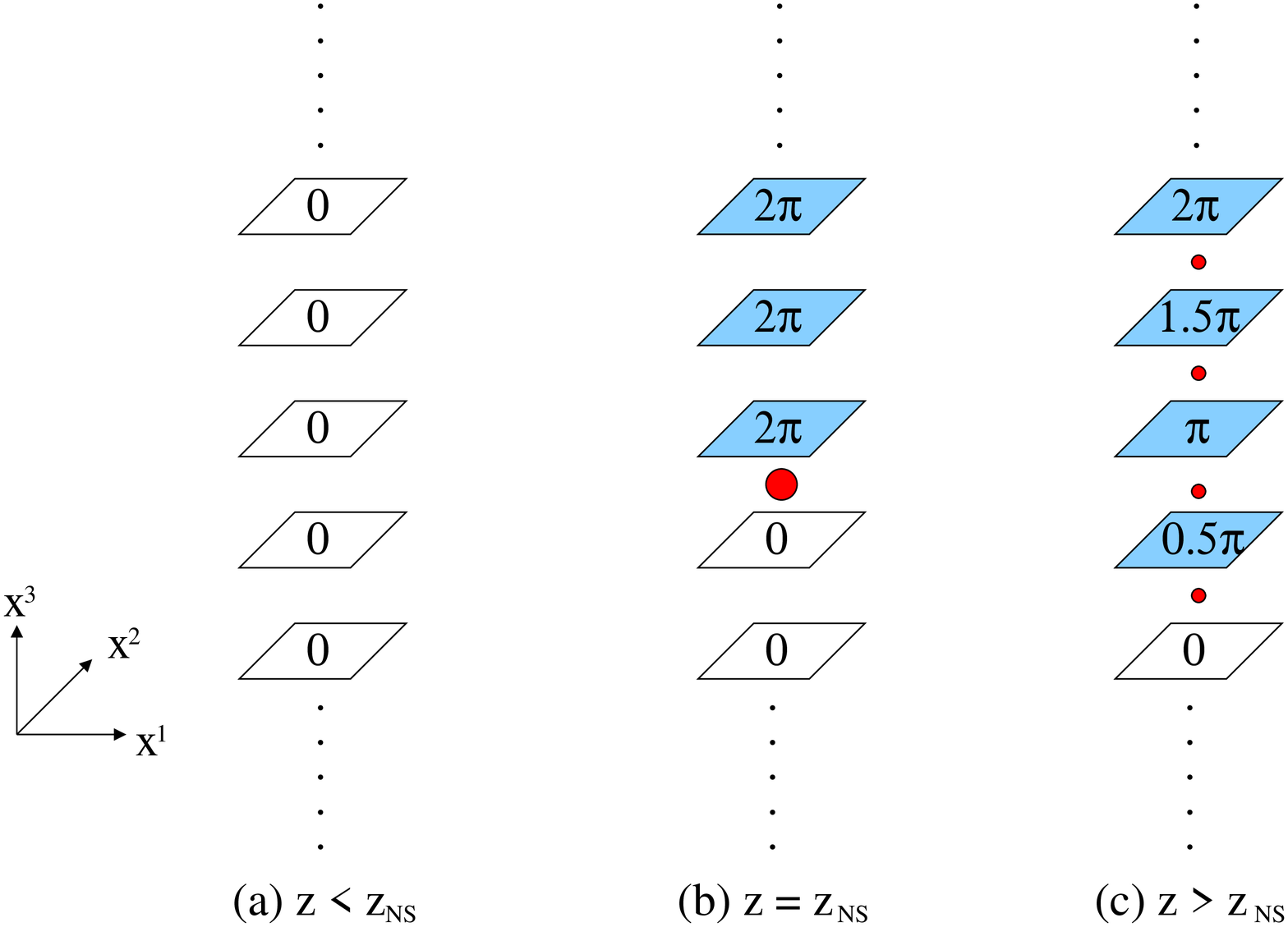}
\caption{
An \NS-brane can be viewed as 
an instanton in $D=3$ 
where one inserts 
a source of $2\pi$ magnetic charge
to the theory at a given $z_{NS}$.
(a) For $z<z_{NS}$, the phases of loop fields are zero.
(b) At $z=z_{NS}$, the phases of loop fields
are $2\pi$ along a semi-infinite line creating
a cube which contains the 3-form flux of $2\pi$ :
$H_{123} = \Delta_1 B_{23} + \Delta_2 B_{31} + \Delta_{3} B_{12} 
= \frac{2 \pi}{a^3} \delta_{x,x_{NS}}$.
(c) For $z>z_{NS}$ the configuration in (b)
is smoothly deformed and  
the flux is smeared over a region which 
contains the net flux $2\pi$.
The size of red dots represents
the amount of 3-form flux contained 
in each cube.
}
\label{fig:NS_instanton}
\end{figure}

\begin{figure}[h!]
\centering
      \includegraphics[height=8cm,width=9cm]{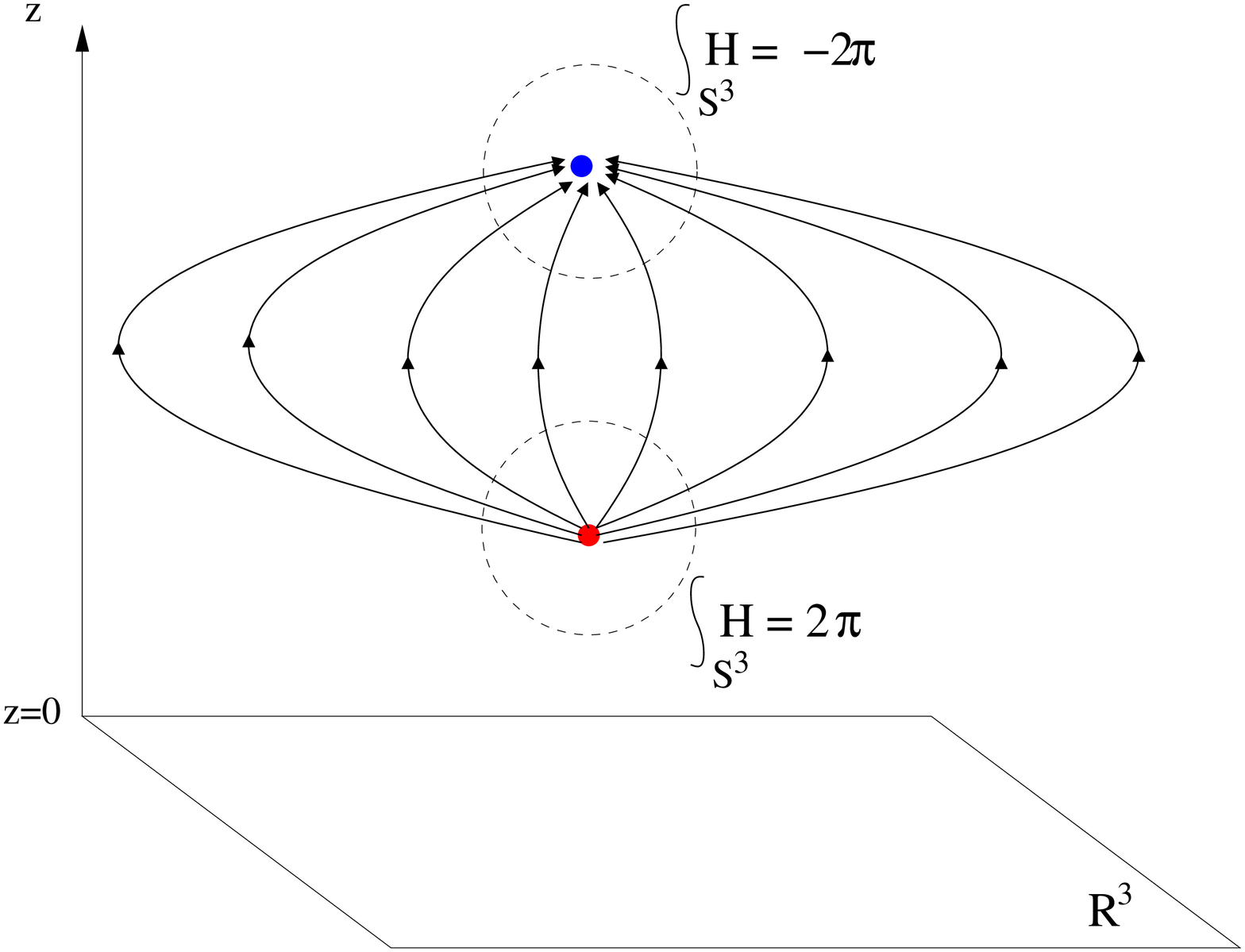}
\caption{
A pair of instanton and anti-instanton
in the four-dimensional bulk for $D=3$.
}
\label{fig:instanton_gas}
\end{figure}

The physical nature of \NS-brane 
can be most easily understood in $D=3$
where \NS-brane is an instanton 
localized at a point in the four-dimensional bulk.
We start with a configuration of 
the loop fields $\phi_\Box$ 
for unit plaquettes with 
\bqa
b_{\Box_{12}}(x,z) & = &
2 \pi 
\delta_{x^1,x^1_{NS}}
\delta_{x^2,x^2_{NS}}
\Theta( x^3 - x^3_{NS} )
 \Theta(z-z_{NS}), 
\eqa
where the phases of the loop fields on 
 $x^1-x^2$ plaquettes 
are $2\pi$ along a semi-infinite line
in the $3$-dimensional space
for $z > z_{NS}$,
and the phases are zero, otherwise.
Since $b_C \equiv b_C + 2\pi$, this configuration
is equivalent to the trivial configuration
where $b_C=0$ everywhere.
However, we can view this configuration 
as a topological defect with
a non-trivial three-form flux on a cube,
$H_{123} = 2 \pi  \delta^3(x-x_{NS}) \Theta(z-z_{NS})$. 
This means that there is a source of magnetic flux
for the two-form gauge field localized at
a point in the bulk,
$d H = 2 \pi \delta^3(x-x_{NS}) \delta(z-z_{NS})$.
Since $2\pi$-magnetic flux is concentrated at one cube, 
the topological defect is trivial.
Now the configuration is deformed smoothly,
$b_C^{'} = b_C + \delta b_C$.
Under a smooth deformation, 
the flux is smeared out over an extended region
in the space,
while the net flux $2\pi$ does not change.
Now the flux is visible by loop fields.
This is illustrated in Fig. \ref{fig:NS_instanton}.
As $z$ is increased further, the flux
can merge back into one cube and disappear
into the vacuum through the inverse process.
This describes a pair of instanton and anti-instanton
as is shown in Fig. \ref{fig:instanton_gas}.
In higher dimensions,
\NS-branes are extended objects.
In $(D+1)$-dimensional bulk,
we can think of Fig. \ref{fig:instanton_gas}
as a configuration in a slice 
at a fixed $x^{5}, ..., x^{D+1}$,
where the \NS-brane is extended along the $(D-3)$
directions.
They can be wrapped into compact objects
as is shown in Fig. \ref{fig:NS},
which in a sense describe bound states
of \NS-brane and anti-\NS-brane.
If the size of the wrapped \NS-branes
become infinite, \NS-brane and anti-\NS branes
become unbound.

The tension of the \NS-brane is proportional to $N^2$
because \NS-brane is a topological defect of 
the two-form gauge field
with the coupling proportional to $1/N^2$.
Here we are using the term `tension' in a loose sense.
In $D>4$, it literally means the tension of \NS-branes.
In $D=4$, it refers to the mass of `\NS-particle'.
In $D=3$, it refers to the action of `\NS-instanton'.
For a sufficiently large $N$, 
we expect that \NS-branes are gapped.
In this case, \NS-branes will be wrapped
into compact objects with a finite size 
in the vacuum.
For a small $N$, the bare tension of  
\NS-brane is small, and quantum fluctuations
can renormalize the tension into a negative value.
Then \NS-branes are condensed,
and extended \NS-branes fill the space in the bulk.
It is also possible that \NS-branes always condense 
for any finite $N$ in low dimensions.
We will discuss the consequences of
different dynamics of \NS-branes
in the following section.

\section{Emergent space and quantum order in Holographic phases }

In this section, we will discuss various phases
that the matrix model can have,
 by focusing 
on the behavior of \NS-branes.
In particular, we will see that
the dynamics of string excitations around a 
saddle-point configuration of the loop fields
is determined by the fate of \NS-branes 
in the bulk.
In order to discuss about this issue systematically,
we first turn to the saddle point equations.

\subsection{Saddle point solution}

The saddle point configuration of loop fields
is determined from the equation of motion.
\bqa
\partial_z \phi_C &=& - L_C \phi_C + 
\frac{1}{M^2} 
\left(
2 F_{ij}[C,C_1] \phi_{C_1}^* \phi_{[C+C_1]_{ij}} 
+
G_{ij}[C_1,C_2] \phi_{C_1} \phi_{C_2} \delta_{(C_1+C_2)_{ij},C}
\right), 
\label{EOM1} \\
- \partial_z \phi_C^* &=& - L_C \phi_C^* + 
\frac{1}{ M^2} 
\left(
F_{ij}[C_1,C_2] \phi_{C_1}^* \phi_{C_2}^* \delta_{[C_1+C_2]_{ij},C}
+ 2 G_{ij}[C_1,C_2] \phi_{(C+C_1)_{ij}}^* \phi_{C_1}  
\right).
\label{EOM2}
\eqa
These equations are supplemented by 
two sets of boundary conditions.
It is more convenient to use the action with
discrete time step $dz$ to isolate boundary fields
from bulk fields.
The UV boundary condition is obtained from Eq. (\ref{S1}),
\bqa
\frac{\partial S_1}{\partial \phi^{(0)*}_C}
= N^2 \left[   
\phi_C^{(0)} + 
\frac{\partial V[\phi_C^{(0)*}]}{\partial \phi^{(0)*}_C}
\right] = 0,
\label{BC_UV}
\eqa
and the IR boundary condition from Eq. (\ref{S7}),
\bqa
\frac{\partial S_7}{\partial \phi^{(R)}_C}
= N^2 \left[   
\phi_C^{(R)*} + 
\frac{\partial V^{'}[\phi_C^{(R)}]}{\partial \phi^{(R)}_C}
\right] = 0.
\label{BC_IR}
\eqa
When the UV potential includes only single-trace operators, 
$V = -J_C \phi_C^{(0)*}$, 
Eq. (\ref{BC_UV}) leads to the standard Dirichlet boundary condition 
for the source field : 
$\phi^{(0)}_C = J_C$.
For more general non-linear UV potential,
it becomes a mixed boundary condition.
This is consistent with the prescription for the UV boundary condition
in the presence of multi-trace deformations
in the standard AdS/CFT dictionary\cite{WITTEN_multi}.
The IR boundary condition is a mixed one
because $V^{'}$ is in general non-linear.
For self-retracting loops,
$V^{'}$ also contains terms 
that are linear in loop fields 
as is shown in the first term in Eq. (\ref{Vp}).
Eq. (\ref{BC_IR}) then implies that $\phi_C^* \neq 0$
at the IR boundary  for self-retracting loops.
As will be shown in the next paragraph,
this means that loop fields for self-retracting loops
have non-zero expectation values at all $z$ in the bulk.
This, in turn, generates non-zero vacuum expectation
values of the source fields $\phi_C$ for self-retracting loops.
As was discussed in Fig. \ref{fig:pair_annihilation},
self-retracting loops can turn into
a loop/anti-loop pair through an interaction.

In general, the saddle point configuration is $z$-dependent,
and $\phi_C(z)$ is not necessarily the
complex conjugate of $\phi_C^*(z)$.
One should treat $\phi_C$ and $\phi_C^*$ 
as two independent fields.
Then the equations of motion can be viewed
as a set of Hamiltonian equations in the phase space 
of $\{ \phi_C, \phi_C^* \}$. 

\begin{figure}[h!]
\centering
      \includegraphics[height=6cm,width=15cm]{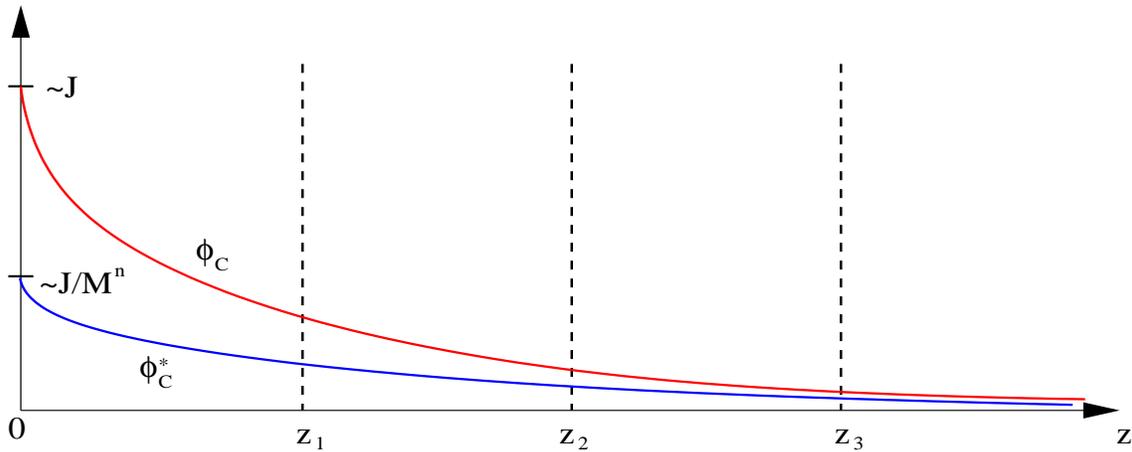}
\caption{
A schematic profile of a loop field $\phi_C$ and
the conjugate field $\phi_C^*$ in a deep confinement phase.
$\phi_C$ and $\phi_C^*$ satisfy the boundary condition
given by Eq. (\ref{BC_IR2}) at any $z$ in the bulk.
}
\label{fig:phi_wall}
\end{figure}

Although Eqs. (\ref{EOM1}) and (\ref{EOM2}) 
are coupled equations for $\phi_C$ and $\phi_C^*$,
one can eliminate $\phi_C^*$ in favor of $\phi_C$.
We first note that the partition function
and all observables including vacuum expectation values
of Wilson loop operators
represented by $\phi_C^*(0)$ are independent of 
how we choose $\beta$ in Eqs. (\ref{Z_main0}) and (\ref{S_bulk0}).
This means that the saddle point solution $\phi_C(z)$ and $\phi_C^*(z)$ for $z < \beta$
is independent of $\beta$.
Since we could have put $\beta$ anywhere, 
$\phi_C$ and $\phi_C^*$ should satisfy 
the IR boundary condition at any $z$,
\bqa
\phi_C^{*}(z) = 
- \frac{\partial V^{'}[\phi_C(z)]}{\partial \phi_C(z)}.
\label{BC_IR2}
\eqa
This is illustrated in Fig. \ref{fig:phi_wall}.
The fact that one can put the IR boundary at any $z$ 
has an interesting implication on the role of the IR boundary.
Usually, one can associate a boundary condition with a physical object located
at the boundary.
However, Eq. (\ref{BC_IR2}) is special in the sense that
an observer at $z < \beta$ can not `feel' the presence of 
a physically identifiable object at $z=\beta$.
Suppose one stops the renormalization group procedure at $z=\beta$
and impose  Eq. (\ref{BC_IR2}) at the IR boundary.
If a UV observer sends a wave toward the IR region,
the reflected wave from the IR region
is exactly the same as the reflected wave one would observe
in the space which is extended to $z=\infty$ without 
an boundary.
In this sense, the IR boundary is not a physical boundary : 
one can always trade the IR boundary with the space
where $z$ is extended to infinity.

Using Eq. (\ref{BC_IR2}), we can write a set of first order differential equations
for the source field only,
\bqa
\partial_z \phi_C &=& - L_C \phi_C + 
\frac{1}{M^2} 
\left(
- 2 F_{ij}[C,C_1] \frac{\partial V^{'}[\phi_{C_1}]}{\partial \phi_{C_1}} \phi_{[C+C_1]_{ij}} 
+
G_{ij}[C_1,C_2] \phi_{C_1} \phi_{C_2} \delta_{(C_1+C_2)_{ij},C}
\right). \nn
\label{EOM1_2}
\eqa
Once $\phi_C(z)$ is solved using the UV boundary condition in Eq. (\ref{BC_UV}), 
the conjugate field is readily determined from Eq. (\ref{BC_IR2}).
One can check that the source and the conjugate field satisfy Eq. (\ref{BC_IR2}) at all $z$
through an explicit calculation perturbatively in $1/M$ (see Appendix C).

It is tempting to interpret Eq. (\ref{EOM1_2}) as the beta
function of the sources for Wilson loop operators.
However, there is an important caveat for this interpretation,
which comes from the fact that this is the saddle point equation
of the quantum theory for dynamical loop fields.
The saddle point equation is expected to be valid 
only when quantum fluctuations are weak
for a sufficiently large $N$.
For a small $N$, one can still have a well-defined beta function
under the usual renormalization group flow\cite{POLCHINSKI84,POLONYI}.
However, the beta function can not be directly identified 
with the saddle point equation of the loop fields if
the saddle point solution becomes unstable
by strong quantum fluctuations.
As we will see in the following sections,
non-perturbative fluctuations 
can invalidate the holographic description
for small $N$.

\subsection{Fluctuations near the saddle point}

Fluctuations near the saddle point configuration $\bar \phi_C(z)$ describes 
dynamical string in the bulk,
\bqa
\phi_C (z) & = & \bar \phi_C (z) +  \chi_C(z) ,
\eqa
where 
$\chi_C$ describes small fluctuations around the saddle point.
We call $\chi_C$ string field
to distinguish it from the loop field $\phi_C$.
The dynamics of string is governed by the action,
\bqa
S_{bulk} & = &  
N^2  \int_0^\beta dz \Bigl[ 
\chi_C^{*} \partial_z \chi_C + L_C \chi_C^{*} \chi_C^{} \nn
&& - \frac{ 1 }{M^2} \Bigl( 
F_{ij}[C_1,C_2] \bar \phi_{[C_1+C_2]_{ij}}(z)  \chi_{C_1}^{*}   \chi_{C_2}^{*}
+ 2 F_{ij}[C_1,C_2]  \bar \phi_{C_1}^{*}(z)  \chi_{[C_1+C_2]_{ij}}    \chi_{C_2}^{*} \nn
&& ~~~~~~ + G_{ij}[C_1,C_2] \bar \phi_{(C_1+C_2)_{ij}}^{*}(z)   \chi_{C_1} \chi_{C_2} 
 + 2 G_{ij}[C_1,C_2] \bar  \phi_{C_1}(z) \chi_{C_2}  \chi_{(C_1+C_2)_{ij}}^{*} \Bigr) \nn
&& - \frac{ 1 }{M^2} \Bigl( 
F_{ij}[C_1,C_2] \chi_{[C_1+C_2]_{ij}}  \chi_{C_1}^{*}   \chi_{C_2}^{*} 
 + G_{ij}[C_1,C_2] \chi_{C_1}^{} \chi_{C_2}^{}  \chi_{(C_1+C_2)_{ij}}^{*} 
\Bigr)
\Bigr].
\label{bulk_chi}
\eqa
Here $\chi_C(z) = \chi_C^r(z) + i \chi_C^i(z)$ is a complex field.
Following the standard method of the steepest descent,
the contours of the real and imaginary parts of the complex fields
are chosen so that 
the real part of the eigenvalues 
of the quadratic action becomes maximum
along the deformed contours\cite{CALLAN_COLEMAN,REINHARDT}.
Note that the string fields $\chi_C$ acquire the hopping term
through non-zero condensates of loop fields.
It also has the terms that describe pair creation/annihilation of two closed strings.

\subsection{Possible phases of the matrix model}

In this section, 
we describe possible states of the matrix model 
using the holographic description.
In particular, we will see that strings have different
dynamics depending on the behavior of \NS-branes.
One observable that is useful in distinguishing different states 
is the correlation function between Wilson loop operators.
In particular, we focus on the correlation function of phase fluctuations of 
Wilson loop operators,
\bqa
F(C,C^{'}) & = & \left< \delta b_C \delta b_{C^{'}} \right>,
\label{WW}
\eqa
where $\delta b_C = b_C - <b_C>$.
In the bulk description, this correlation function corresponds to a two-point string-string correlation function.
This object is of particular interest 
because the string state that corresponds to the phase mode describes
the two-form gauge field in the bulk.

\begin{figure}[h!]
\centering
      \includegraphics[height=12cm,width=13cm]{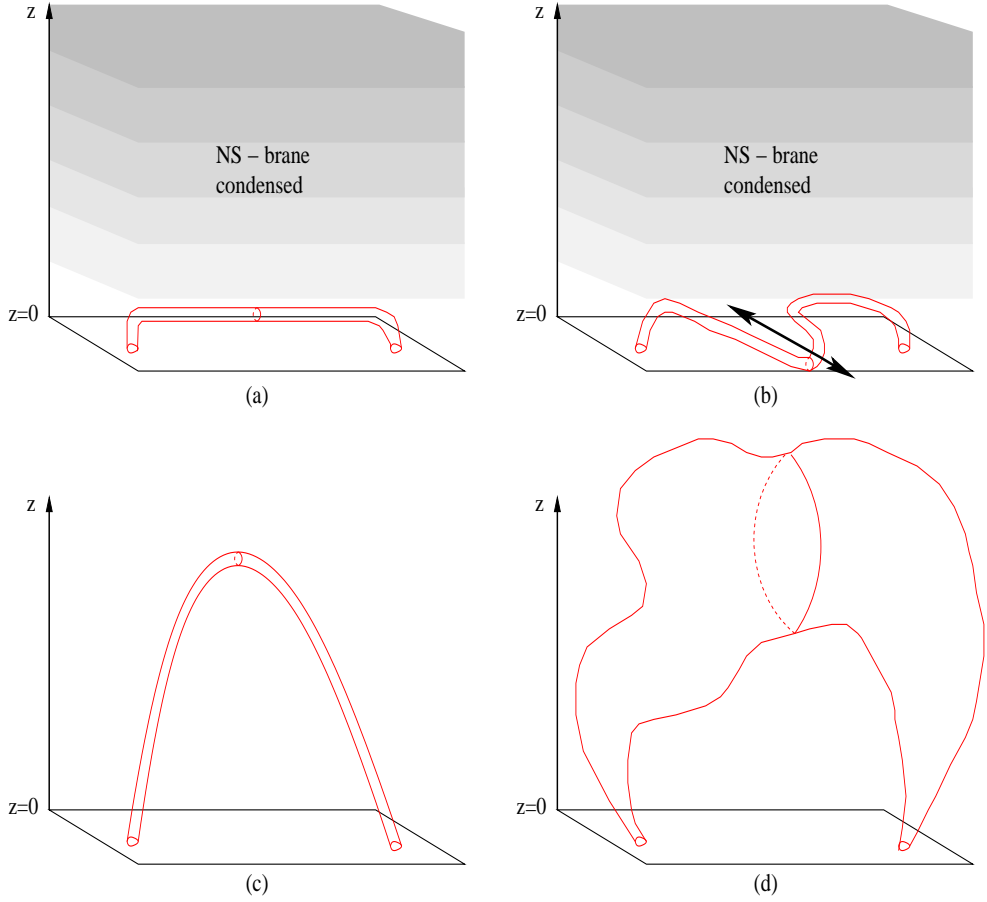}
\caption{
The Wilson loop correlation function in different phases.
(a) In the confinement phase, \NS-branes are condensed in the bulk, and the string emitted at the boundary stays near the boundary. Due to weak fluctuations of string world sheet,
the string propagates in a straight path.
(b) In the non-holographic critical phase, \NS-branes remain condensed in the IR region. 
However, large amplitudes of loop fields in the UV region
make the string world sheet highly fluctuating along the $D$ dimensions. As a result, the correlation 
function decays algebraically.
(c) In the holographic critical phase, loop fields acquire
finite expectation values everywhere in the bulk, and
\NS-brane is gapped.
Strings can propagate deep inside the bulk,
and the correlation function shows a power-law behavior
with a scaling dimension determined by the mass of the string. 
(d) In the deconfinement phase, loop fields with infinitely large loops acquire non-zero expectation value in the bulk.
This causes highly non-local fluctuations
of the world sheet of a string in the bulk.
}
\label{fig:WW_correlation}
\end{figure}

\subsubsection{Confinement phase}

For non-self-retracting loops, 
$V^{'}$ is quadratic or of higher order 
in the loop fields.
For small $\phi_C(0)$, 
the first term on the right hand side of Eq. (\ref{EOM1_2})
dominates.
As a result, $\phi_C(z)$ decays exponentially in $z$.
Because larger loops for which sources are not turned on at the UV boundary
are generated out of many small loops,
amplitudes with larger loops decay exponentially
with the area enclosed by the loop.
This corresponds to the confinement phase of the matrix model.
In the confinement phase, one can define 
a cross-over scale $z^*$
beyond which loop fields have negligible amplitudes.

Now let us consider dynamics of strings in the IR ($z>z^*$)
and the UV ($z<z^*$) regions separately.
In the deep IR region, amplitudes of 
non-self-retracting loop fields are exponentially small. 
This means that loops that are emitted from the UV boundary
rarely reach the IR region.
In this region, phase fluctuations of loop fields do not have a stiffness, 
which leads to a condensation of \NS-branes.
As a result, strings are subject to the strongly fluctuating two-form gauge field.
In this region, strings are confined,
and the two-form gauge field is gapped\cite{POLYAKOV87,REY1991}.
For a more systematic discussion on possible phases of anti-symmetric fields, see Ref. \cite{QUEVEDO}.
In the deep IR region, strings can not propagate by themselves;
only charge neutral bound states of string and anti-string 
can propagate.
On the other hand, loop fields have significant amplitudes in the UV region.
The source field $J_\CCC$ plays the role of a 
symmetry-breaking field
at the UV boundary.
As a result, phase fluctuations of loop fields are small, 
and \NS-branes are suppressed in the UV region.
Because loop fields are coherent near the UV boundary,
strings are deconfined in this region.
There is a domain wall that separates
the IR region with condensed \NS-brane
and the UV region without \NS-brane.

In the confinement phase, a string 
that is emitted from the boundary can not penetrate through 
the wall of condensed \NS-brane.
Therefore it stays within the UV region.
Deep inside the confinement phase with small $J_\CCC$, 
the condensates of loop fields are small.
As a result, the hopping amplitudes of strings are small,
and fluctuations of string world sheet is small.
For the correlation function in Eq. (\ref{WW}),
the strings inserted at the UV boundary
are connected through a minimum number of hoppings,
forming a straight path
as is shown in Fig. \ref{fig:WW_correlation} (a).
This leads to an exponentially decaying correlation 
function for the Wilson loop operators.
In the confinement phase, the bulk geometry ends at a finite scale 
due to the proliferation of \NS-branes.
This is reminiscent of the idea that geometry can
get truncated by tachyon condensation\cite{HS2006}.

As $J_\CCC$ is dialed up,
amplitudes of loop fields in the bulk increase.
Accordingly the cross-over scale $z^*$ increases.
At the same time, 
fluctuations of string world sheet increase
as the amplitudes of loop fields become larger.
Suppose the system becomes critical 
either by fine tuning or dynamical tuning.
In the case of fine tuning, 
one may have to tune more 
than one microscopic parameters
to reach a critical point.
For the following discussions 
which focus on physical properties 
of the critical states,
it is not important whether 
those states are realized 
as phases or critical points.
So we will use the term `critical phase' 
in a broad sense to include 
not only critical phases realized
within a finite region in the parameter space
of a microscopic model
but also critical states realized at critical points
by fine tuning.
Logically, there exist 
at least two different scenarios
via a criticality is achieved.
In the first scenario, \NS-branes remain condensed in the 
IR region with a finite $z^*$.
But loop fields acquire large amplitudes in the UV region
so that strings are delocalized along the $D$-directions, 
mediating critical correlations between operators 
inserted on the UV boundary.
In the second scenario, the cross-over scale diverges
and \NS-branes are suppressed out in the bulk.
In this case, strings can propagate deep inside the bulk.
In the following, 
we will discuss the two scenarios in more detail.

\subsubsection{Non-holographic critical phase }

Here we discuss the first case
which is likely to be realized when $N$ is small.
For a small $N$, \NS-branes are `light'. 
Even though loop fields have finite amplitudes in the bulk,
quantum fluctuations may destabilize the saddle-point solution.
Indeed this is what always occurs in the pure 2-form
gauge theory in the flat four dimensional space\cite{POLYAKOV87,ORLAND82,PEARSON}.
If the mass or tension of \NS-brane becomes negative,
\NS-branes are condensed.
Once \NS-branes are proliferated, 
the two-form gauge field acquires a mass gap, 
and strings are confined in the bulk,
as is the case in the confinement phase.
The difference from the confinement phase is that
the boundary theory is critical.
The boundary theory can be critical
although strings are confined deep 
inside the bulk because critical correlation 
between boundary fields can be mediated 
by strings that propagate near the UV region.
It is noted that 
`confinement' in the boundary matrix model
and `confinement' of strings in the bulk
are not the same thing.
In this phase, strings are confined in the bulk,
but the matrix model is not in the confinement phase.
We call this {\it non-holographic critical phase}.

In this critical phase, 
a string emitted from the UV boundary
no longer takes the straight path because
of large amplitudes of background loop fields
in the UV region.
Rather, the world sheet of string strongly fluctuates, 
and the correlation function can decay in a power law
because of delocalized strings.
However, strings are still localized
within the UV region along the $z$ direction
due to the condensation of \NS-branes in the IR region.
The fluctuations of the world sheets of strings
is predominantly along the $D$ space dimensions
as is illustrated in Fig. \ref{fig:WW_correlation} (b).

\subsubsection{Holographic critical phase I : deconfined closed string }

The second scenario is qualitatively different from the previous one.
In this phase, loop fields develop non-zero amplitudes
in the IR region.
For a sufficiently large $N$, \NS-branes are suppressed,
and the saddle-point solution is stable
against non-perturbative fluctuations of the two-form gauge field.
If \NS-branes are suppressed,
the compactness of the gauge field is unimportant at long distances.
Strings are deconfined in the bulk because
loop fields provide a background  
in which strings can propagate coherently.
Note that strings are still coupled 
with the dynamical two-form gauge field,
but the gauge field is no longer confining
in this phase.
This is analogous to the Coulomb phase 
discussed in Sec. II.
Thanks to the coherent background loop fields,
closed strings can explore the extended space in the bulk.
The bulk space is not a gauge invariant object.
However, it assumes a `classical identity' 
in the large $N$ limit 
where fluctuations of loop fields are suppressed.
In this sense, the bulk space emerges
in the holographic phase, but not
in the non-holographic phase.

Because of the source that explicitly breaks
the gauge symmetry,
the transformation 
$B_{MN} \rightarrow B_{MN}
+ \partial_M \theta_N - \partial_N \theta_M$,
is not a symmetry at the UV boundary.
As a result, the $U(1)$ mode $\theta_M$ becomes
a physical mode.
This means that there is a $U(1)$ gauge field localized 
at the UV boundary in the dual theory.
This $U(1)$ mode originates
from the Abelian component of the $U(N)$ matrix theory.
We identify this as the singleton mode localized at the boundary\cite{WITTEN,Aharony_Witten98,Maldacena2001}.

It is noted that the symmetry breaking source $J_\CCC$ at the UV boundary does not necessarily 
open up a gap for the phase modes of loop fields in the bulk.
This is because the symmetry breaking source is only at the UV boundary,
but not in the bulk.
This is analogous to the case where one applies a symmetry breaking field
at the boundary of a system 
where a global symmetry is spontaneously broken in the bulk.
Although the boundary field determines the direction of the symmetry breaking
in the whole system, 
the Goldstone mode in the bulk survives in the thermodynamic limit.

The dynamics of the $D$-dimensional matrix model
in the long distance limit is governed by strings
that propagate in the $(D+1)$-dimensional space.
We call this phase {\it holographic phase}.
The hallmarks of the holographic phase are
deconfined strings that propagate
in the bulk with the extra dimension
and the emergence of the Bianchi identity $dH=0$
for the two-form gauge field
in the long distance limit.
It is emphasized that these features
are not protected by any symmetry 
of the original matrix model.
They are dynamical properties 
which emerge only in the holographic phase.
The quantum order present in the holographic phase 
is analogous to the quantum order associated 
with the emergent Bianchi identity
in the fractionalized phase of the
slave-particle theory as discussed
in Sec. II.

In the holographic critical phase,
strings can propagate deep inside the bulk 
as is shown in Fig. \ref{fig:WW_correlation} (c).
The correlation function shows a power-law decay
through a classical trajectory that is extended to the bulk.
Because of the gauge symmetry and the non-trivial quantum order,
we expect that the scaling dimension of the phase mode of Wilson 
loop operators will be protected accordingly.
To determine the scaling dimension,
one has to first solve the loop equations in the bulk
and find the string Green's function 
in the background determined by the loop fields.
We defer an explicit calculation for future studies.

In the four-dimensional ${\cal N}=4$ super Yang-Mills theory,
the phase mode of Wilson loop operators 
has the scaling dimension $6$\cite{DAS98}.
The reason why it is not $2$,
which is the expected scaling dimension 
for massless two-form fields in $D=4$\cite{WITTEN},
is that a Chern-Simons coupling 
generates a mass for the two-form gauge
field through the mixing with
the Ramond-Ramond fields.
However, the scaling dimension is still protected 
from acquiring a large quantum correction
in the large $N$ limit.
Such protection of scaling dimension is often
attributed to supersymmetry.
However, the non-trivial quantum order 
will protect the scaling dimension of the phase mode
from acquiring a large quantum correction
even in non-supersymmetric holographic critical phases 
in the large $N$ limit.
This is true whether or not 
the two-form gauge field becomes massive 
through a Chern-Simons coupling 
with Ramond-Ramond fields.
This is because the string theory becomes classical 
in the large $N$ limit, 
and the coefficient of the Chern-Simons term 
is quantized.
Therefore the mass of the two-form gauge field
can not become large even when other string modes
become very massive 
at strong coupling 
of the boundary matrix model.
This, in turn, implies that the scaling dimension remains small
for the phase mode of Wilson loop operators.
This is in sharp contrast to the 
non-holographic critical phase
where it is expected that the operator generally
receives a large quantum correction at strong coupling.

\subsubsection{Deconfinement phase}

Strictly speaking, the critical phases discussed in 
the previous two sections are kinds of deconfinement phases.
Here we use the term 'deconfinement phase' in a narrower meaning,
that is, free theory in the IR limit. 
If the sources at the boundary are very large,
the second and third terms in Eq. (\ref{EOM1_2}) dominate,
and the source fields will grow as $z$ increases
for $D>4$ in which case the
boundary matrix model is expected to 
flow into IR free gauge theory
in the weak coupling (large $J_\CCC$) limit.
As the amplitudes of the loop fields become larger,
large loops are generated through the joining processes.
As a result, loop fields with all sizes
are condensed in the bulk.
Then strings propagating in the bulk become highly 
non-local because strings can hop from one configuration
into another configuration which is very different from 
the initial one.
This is a string condensed phase.
In this phase, the two-form gauge field acquires a mass
and \NS-brane is confined 
due to the Higgs mechanism\cite{REY89,YI99}.

In the deconfinement phase, 
a string emitted from the boundary
becomes very large in the bulk
and lose its identity as a closed string.
The critical fluctuations are mediated by
highly non-local fluctuations in the bulk.
In this phase, the locality is lost in the bulk.

The deconfinement phase can be viewed as
an extreme limit of the holographic critical phase discussed
in the previous section.
Even in the holographic critical phase,
some loop fields are condensed in the bulk,
as is the case in the deconfinement phase.
The difference is that only small loops are condensed
in the holographic critical phase while
loops with all sizes are condensed in the deconfinement phase.
Note that condensations of small loops do not generate a mass gap
for the two-form gauge field.
This is because closed strings with finite sizes 
as point-like particles
are coupled only with the field strength tensor
of the two-form gauge field.
In certain models, one can in principle change 
microscopic parameters to  tune the size of condensed loops,
smoothly interpolating between 
the holographic critical phase
and the deconfinement phase.
This basically controls the size of strings in the bulk.
The ${\cal N}=4$ SU(N) gauge theory in four dimensions 
is believed to be
in this class : for a sufficiently large $N$,
one can smoothly tune the 't Hooft coupling $\lambda$
from a large value to zero without going through a phase transition.
The one parameter family of the critical theories form a line of fixed points.
Here, $\lambda=0$ is a special point where
the size of string diverges.
In non-supersymmetric theories, it is expected to be harder to 
stabilize a theory at an arbitrary gauge coupling.
Most likely, we expect that the holographic critical 
phase will arise
as a multi-critical point between
the confinement phase
and the deconfinement phase 
for a sufficiently large $N$.

\subsection{A mean-field description}

Some features of the phases discussed in the previous section
can be easily understood if we focus on 
a subspace within the space of 
loop fields $\{ \phi_C \}$.
We focus on the mean-field Ansatz\cite{REY89,YI99}
where a loop field is represented by
a product of link fields along the loop,
\bqa
\phi_C(z) = \prod_{(i,i+\mu) \in C} \xi_{\mu}(i,z).
\label{loop_decomposition}
\eqa
Here $\xi_{\mu}(i,z) = \xi_{-\mu}^*(i+\mu,z)$ is a complex scalar field defined
on the link $i,i+\mu$.
This is a huge simplification where
we reduce the space spanned by 
functions defined on loop space 
into the space spanned by functions 
defined on the links.
Under the gauge transformation,
the link variables transform as
\bqa
\xi_{\mu}(i,z) \rightarrow e^{i \theta_{\mu}(i,z)} \xi_{\mu}(i,z).
\label{u2}
\eqa
Therefore, these link fields should be charged with respect
to the two-form gauge field.
A minimal action for the link field 
that has the same symmetry as the original loop model
is an Abelian-Higgs model\cite{REY89} for the link field,
\bqa
S_{bulk} & = & 
-t \sum_\Box 
\left[ \xi_M(i) \xi_N(i+M) \xi_M^*(i+N) \xi_N^*(i) e^{-i B_{MN}(i)} + c.c. \right]
+ \sum_{i,M} V( | \xi_M(i)|^2 ) \nn
&& - \frac{1}{g_{KR}^2} 
\sum_{\mbox{cubes}} \cos( \Delta_L B_{MN} + \Delta_{M} B_{NL} + \Delta_N B_{LM}).   
\eqa
Here we discretize the $z$ direction, 
and the action is written in a $(D+1)$-dimensional lattice.
The box represents sum over all plaquettes including temporal plaquettes.
The link fields $\xi_z$ along the temporal directions can be viewed
as an auxiliary field that is introduced to keep the two-form gauge symmetry.
One can use a continuum description as well\cite{YI99}.

The phase structure of this model is very similar 
to the one for the Abelian-Higgs model for 
scalar fields discussed in Sec. II.
If \NS-branes are condensed in the bulk,
the link fields are confined.
This is what happens in 
the confinement phase 
and the non-holographic critical phase
discussed in the previous section.
The bulk physics alone can not distinguish
the confinement phase and the non-holographic critical phase.

If the link fields are condensed,
the two-form gauge field acquires a mass
due to the Higgs mechanism\cite{REY89,YI99}.
Note that loop fields 
with arbitrarily large size acquires expectation values
in the Higgs phase 
because loop fields are just products of link fields.
This corresponds to the deconfinement phase 
of the boundary matrix model.

If the link field is gapped 
and the two-form gauge coupling is small,
the theory can be in the Coulomb phase. 
In this phase, closed strings are deconfined
and the two-form gauge field
arises as a light mode in the bulk.
This corresponds to the holographic
critical phase.
Note that the loop fields can have finite expectation values
even though link fields are gapped in this phase.

\subsection{Holographic critical phase II : deconfined open string}

\begin{figure}[h!]
\centering
      \includegraphics[height=3.5cm,width=11cm]{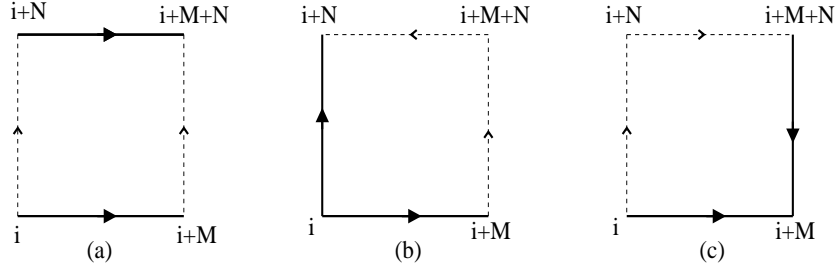}
\caption{
The figures in (a), (b) and (c) show
a `particle' denoted as a thick line 
defined on the link $<i,i+M>$ hops 
to the link $<i+N,i+M+N>$,
 $<i,i+N>$ and $<i+M+N,i+M>$, respectively.
The dashed lines represent the paths
along which the end points of the link follow
to which the U(1) gauge field is electrically coupled.
}
\label{fig:link_hopping}
\end{figure}

The mean field description discussed 
in the previous section
allows one to understand a yet another new phase 
of the matrix model.
To see this, we first note that the decomposition
in Eq. (\ref{loop_decomposition}) has a U(1) gauge redundancy,
\bqa
\xi_M(i) \rightarrow e^{i (\gamma_{i+M} - \gamma_i)} \xi_M (i),
\label{u1}
\eqa
where $\gamma_i$ is a U(1) phase defined on each site
on the bulk space.
Because of this U(1) redundancy, 
the link field can not have a quadratic hopping term.
This is similar to the U(1) gauge redundancy present
in the slave-particle theory discussed in Sec. II.
One can decouple the quartic term for the link fields 
using the Hubbard-Stratonovich transformation.
The resulting action should include a dynamical compact 
U(1) gauge field,
\bqa
S_{bulk}^{'} & = & 
-t \sum_{i, M\neq N} 
\left[ \xi_M^*(i+N) \xi_M(i) e^{-i \left[ B_{MN}(i) - A_N(i+M) + A_N(i) \right]  } + c.c. \right] \nn
&& -t \sum_{i, M\neq N} 
\left[ \xi_N^*(i) \xi_M(i) e^{-i \left[ B_{MN}(i) - A_N(i+M) + A_M(i+N) \right]  } + c.c. \right] \nn
&& -t \sum_{i, M\neq N} 
\left[ \xi_N(i+M) \xi_M(i) e^{-i \left[ B_{MN}(i) + A_M(i+N) + A_N(i) \right]  } + c.c. \right] \nn
&& + \sum_{i,M} V( | \xi_M(i)|^2 ) 
 - \frac{1}{g_{KR}^2} 
\sum_{\mbox{cubes}} \cos( \Delta_L B_{MN} + \Delta_{M} B_{NL} + \Delta_N B_{LM}).
\label{S10}
\eqa
Under the gauge transformation in Eq. (\ref{u1}), 
the U(1) gauge field transforms as usual, $A_M(i) \rightarrow A_M(i) + \gamma_{i+M} - \gamma_i$,
and the two-form gauge field is invariant.
Eq. (\ref{u1}) implies that the link field $\xi_M(i)$
carries U(1) charge $+1$ on one end at site $i+M$
and charge $-1$ on the other end at site $i$.
Under the two-form gauge transformation in Eq. (\ref{u2}),
the U(1) gauge field transforms as
\bqa
A_M(i) \rightarrow A_M(i) + \theta_M(i).
\eqa
The first term in Eq. (\ref{S10}) describes
a link parallel to the direction $M$ 
hops along the direction $N$
as is shown in Fig. \ref{fig:link_hopping} (a),
and the second and third terms describe hoppings 
where the direction of a link changes 
as in Fig. \ref{fig:link_hopping} (b) and (c).

Although the bare coupling for the U(1) gauge field
is infinite, the kinetic term will be generated 
once high energy fluctuations of the link fields 
are integrated out, renormalizing the gauge coupling
to a finite value.
Suppose we are in the holographic critical phase
where the link fields are gapped
and the two-form gauge field is in the deconfinement phase.
The U(1) gauge field can be either in the confinement phase
or in the deconfinement phase.
If the U(1) gauge field is confining, 
all open links are joined with each other 
to form closed loops 
because there is a linearly confining force between open ends.
In this phase, only closed strings are allowed.
This is the holographic critical phase I 
discussed in Sec. VII.C.3.
On the other hand, if the U(1) gauge field is 
in the deconfinement phase,
closed strings can get fractionalized into open strings,
and open strings arise as deconfined excitations.
Closed strings can still exist as a bound state of open strings.
In this phase, there is an emergent U(1) gauge field
in addition to the two-form gauge field.
The two-form gauge field is coupled with the world-sheet 
of strings and the U(1) gauge field is coupled
with boundaries of open strings.
Here the gapless U(1) gauge field is protected by 
a quantum order associated with 
the emergent Bianchi identity $dF=0$
for the U(1) gauge field in the bulk.
Open strings are fractionalized 
excitations of closed strings.

\begin{figure}[h!]
\centering
      \includegraphics[height=4cm,width=10cm]{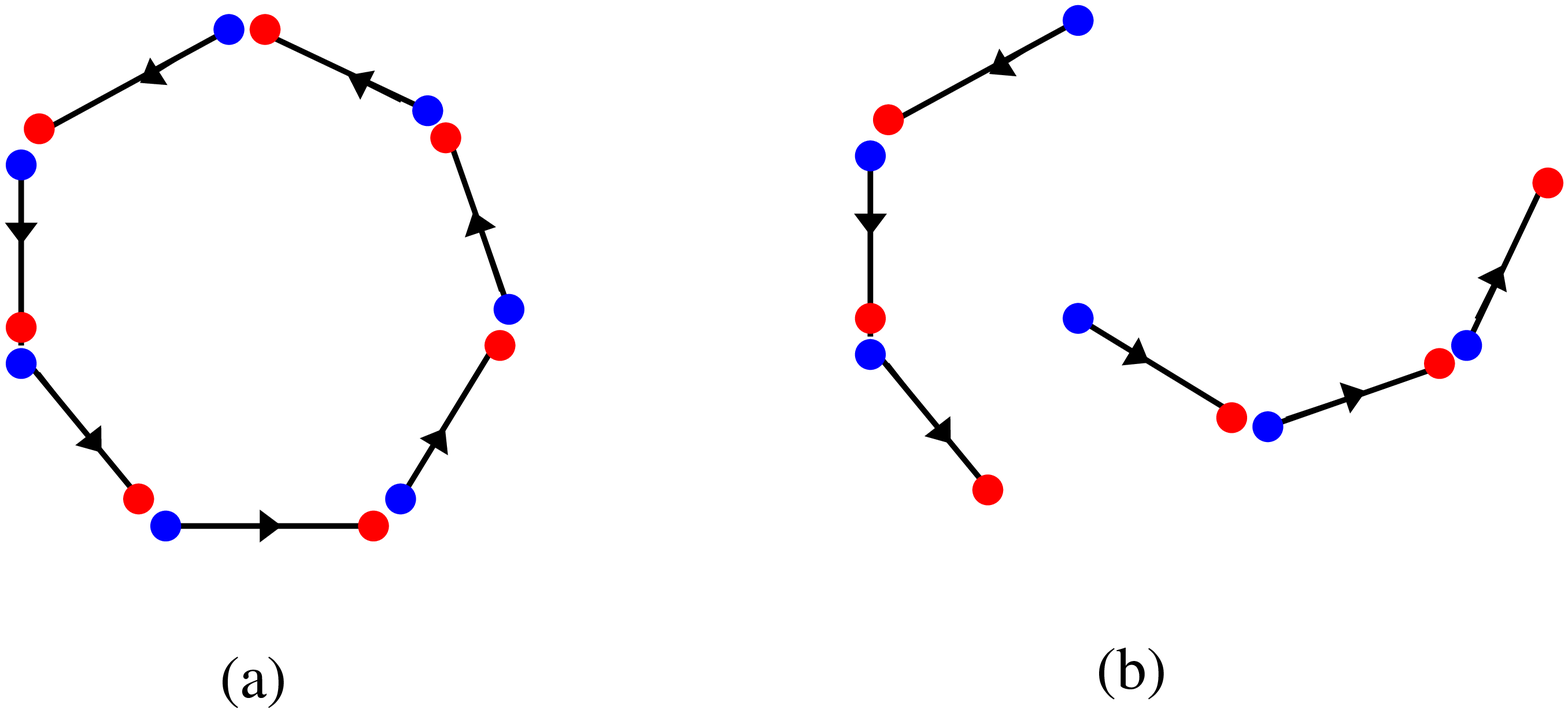}
\caption{
(a) In the confinement phase for the one-form gauge field in the bulk,
link fields are always joined to form closed loops,
and there exist only closed strings as finite energy states.
(b) In the deconfinement phase, closed strings can be
fractionalized into open strings.
The end points of open strings carry gauge charge for
the emergent gauge field in the bulk.
}
\label{fig:conf_deconf_open}
\end{figure}

The way open strings and the U(1) gauge field 
arise as collective excitations 
of closed string fields is very similar to the
way slave-particles emerge as fractionalized 
excitations along with the emergent U(1) gauge boson
in the slave-particle theory discussed in Sec. II.
One can go one step further to obtain a 
different gauge group for open strings.
For this we introduce a larger gauge redundancy in Eq. (\ref{loop_decomposition}),
\bqa
\phi_C(z) = \tr \left[ \prod_{(i,i+\mu) \in C} \Xi_{\mu}(i,z) \right],
\label{loop_decomposition2}
\eqa
where $\Xi_{\mu}(i,z) = \Xi_{-\mu}^\dagger(i+\mu,z)$ is a $\td N \times \td N$ complex matrix field
defined on the link $<i,i+\mu>$ in the bulk. 
This decomposition has the $U(\td N)$ gauge redundancy,
\bqa
\Xi_{\mu}(i,z) \rightarrow V_i^\dagger(z) \Xi_{\mu}(i,z) V_{i+\mu}(z).
\eqa
Therefore the link fields now have to be coupled with
a dynamical $U(\td N)$ gauge field in the bulk.
\bqa
S_{bulk}^{''} & = & 
-t  \sum_{i, M\neq N} 
\tr \left[ \Xi_M^\dagger(i+N) U_{i+N,i} \Xi_M(i) U_{i+M,i+M+N} 
 e^{-i B_{MN}(i) } + c.c. \right] \nn
&& -t  \sum_{i, M\neq N} 
\tr \left[ \Xi_N^\dagger(i) \Xi_M(i) U_{i+M,i+M+N} U_{i+M+N,i+N} 
e^{-i B_{MN}(i) } + c.c. \right] \nn
&& -t  \sum_{i, M\neq N} 
\tr \left[ \Xi_N(i+M) \Xi_M(i) U_{i+M+N,i+N} U_{i+N,i} 
e^{-i B_{MN}(i) } + c.c. \right] \nn
&& + \sum_{i,M} V( | \Xi_M(i)|^2 ) 
 - \frac{1}{g_{KR}^2} 
\sum_{\mbox{cubes}} \cos( \Delta_L B_{MN} + \Delta_{M} B_{NL} + \Delta_N B_{LM}).
\label{S11}
\eqa
Here $U_{i,i+M} = e^{ i \tau^a A^a_M(i) }$,
and $A^a_M(i)$ is $U(\td N)$ gauge field defined on 
the link $<i,i+M>$.
It is noted that the $U(\td N)$ gauge field that emerges
in the bulk is different 
from the original $U(N)$ gauge field
of the boundary matrix model.
In this description, end points of the link fields
carry fundamental and anti-fundamental charges
for the dynamical $U(\td N)$ gauge field.
If the $U(\td N)$ gauge field in the bulk is in 
the deconfinement phase,
open strings with the $U(\td N)$  Chan-Paton factor
emerge as collective excitations of the closed
string fields.

Although one can choose a description with an arbitrary gauge redundancy, the gauge group is determined by dynamics in the end.
Holographic theories with different gauge groups
for open strings in the bulk describe different states
of the matrix model.
In the T-dual description,
configurations with background gauge fields 
describe D-branes.
It is interesting to note that D-branes can emerge as non-perturbative excitations
in the closed string field theory.

\section{Discussion}

\begin{center}
 \begin{table}[h]
 \caption{}
   \begin{tabular}{| c | c | c | c| c|  }
     \hline
                & closed string & open string &  NS-brane & bulk excitations  \\ \hline \hline
 Confinement phase  & confined & confined & condensed &  $\times$ \\ \hline
Non-holographic & confined  & confined & condensed  & $\times$ \\    
 critical phase  & & & & \\    \hline
Holographic & deconfined & confined & gapped &  closed string ($B_{MN}$), \\    
critical phase I & & & & NS-brane   \\    \hline
Holographic & deconfined & deconfined & gapped & closed \& open string ($B_{MN}$, $A_M$), \\
critical phase II & & & & NS-brane, D-brane \\    \hline
Deconfinement (IR free) phase & condensed & condensed & confined & non-local string  \\    \hline
   \end{tabular}
  \end{table}
 \end{center}

In summary, we showed that a $D$-dimensional
gauged matrix model can be mapped into
a closed string field theory in $(D+1)$-dimensional space.
The string field in the bulk is coupled with a 
compact two-form gauge field which is also
a part of the string field.
Holographic states with
deconfined string in the bulk are stable 
only when topological defects 
for the two-form gauge field are suppressed
in the bulk, which is likely to be realized
for a sufficiently large $N$.
The holographic states are in different universality
classes from the non-holographic states 
where strings are confined in the bulk
due to condensed topological defects.
We also discussed a holographic critical state
where closed strings get fractionalized into open strings.
In this state, there are both closed and open strings 
along with the two-form and one-form gauge fields
in the bulk.
The non-trivial quantum order present in 
the holographic phases is responsible for 
the existence of operators whose scaling 
dimensions are protected, which otherwise
would have received a large quantum correction
at strong coupling.
The possible phases of the matrix model
are summarized in Table. II.

Although many structures on the holographic description
have been learned from general considerations,
it is desirable to obtain explicit
solutions to the saddle point equation.
In principle, one has to solve a set of infinitely many
coupled differential equations.
In the future, it will be interesting 
to simplify these equations
by focusing on light modes.
Finally, we close with discussions
on a speculative phase diagram of the matrix model,
a world-sheet description of deconfined strings
in the holographic phases,
and a continuum limit.

\subsection{A schematic phase diagram}

\begin{figure}[h!]
\centering
      \includegraphics[height=6cm,width=8cm]{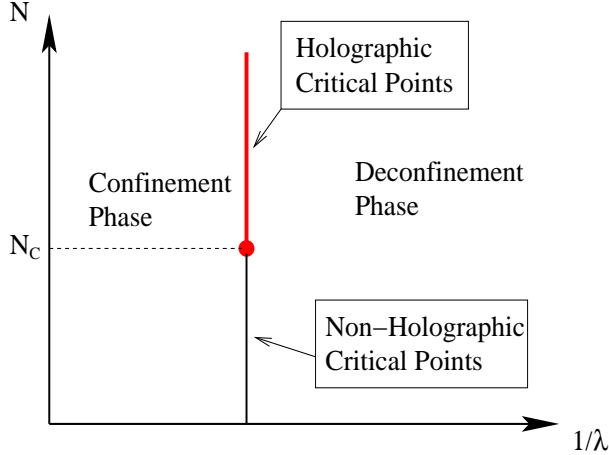}
\caption{
A proposed phase diagram for the pure bosonic gauged matrix model in $D>4$.
Here $\lambda$ represents a set of 
't Hooft couplings associated with various
Wilson loops in the matrix model.
Moving along the direction of $\lambda$ 
in this figure
may mean tuning more than one microscopic parameters
at the same time.
The critical points, in general, represent multi-critical points.
Therefore the shape of the phase boundary in the figure should 
not be taken seriously.
The actual phase boundary is not likely to be a straight line
in the multi-dimensional space of microscopic couplings.
In the strong coupling limit (large $\lambda$),
the theory is in the confinement phase with the
exponentially decaying Wilson loop correlation function.
As the coupling becomes smaller, the model goes through 
a phase transition to the IR free deconfinement phase.
For small $N$, \NS-branes remain condensed at the critical 
point.
Strings are confined, and 
the two-form gauge field is massive in the bulk.
Critical correlations are mediated by 
the $D$-dimensional fluctuations of strings 
near the UV boundary.
Scaling dimensions generally receive a large
quantum correction at strong couplings.
This is the non-holographic critical point.
For large $N$, \NS-brane is gapped
and the two-form gauge field remains light
even at strong coupling.
Low energy physics is governed by deconfined 
strings that propagate deep inside the bulk.
In this holographic critical point,
critical correlations of the matrix model
is mediated by $(D+1)$-dimensional fluctuations of strings.
The scaling dimension of the operator associated 
with the two-form gauge field is protected from 
acquiring a large quantum correction 
due to the quantum order
associated with the dynamical suppression of \NS-branes.
The holographic critical point
can be either in the state with only closed strings,
or in the state with both closed and open strings.
}
\label{fig:phase_diagram}
\end{figure}

It may be difficult to find a specific microscopic model which 
realizes each phase discussed in this paper.
However, one may still guess a possible phase diagram. 
For $D \leq 4$, it is believed that the present matrix model
is always in the confinement phase.
In these low dimensional cases, 
one may have to introduce more 
degrees of freedom (fermions or fundamental matters)
to stabilize critical phases.
Here we focus on the pure bosonic matrix model in $D>4$.
In Fig. \ref{fig:phase_diagram},
we show a speculative phase diagram.
In the strong coupling limit, the matrix model is in the confinement phase.
As the gauge coupling is weakened, 
there is a phase transition 
into the IR free deconfinement phase. 
If the phase transition is continuous,
there can be two different universality classes
for the critical points.
For small $N$, \NS-branes are condensed and
strings are confined in the bulk.
At this non-holographic critical point,
the scaling dimension of Wilson loop 
operators generally receive a large quantum correction.
For $N$ greater than a critical value,
\NS-branes are suppressed, 
and strings are deconfined in the bulk.
The quantum order protects the two-form gauge field
from acquiring a large mass in the large $N$ limit, 
which in turn protects the scaling dimension of 
the phase fluctuations of Wilson loop operators
even at large 't Hooft couplings.

As was discussed in Sec. VII. E, 
there exist two different kinds of holographic critical points.
In the first case, there are only closed string excitations in the bulk.
In the second case, there are both closed and open string excitations along with the emergent gauge field 
and the two-form gauge field.
We expect that it is easier to stabilize the 
state with both closed and open strings 
for $N$'s that are large but not too large.
In the large $N$ limit, closed strings are free,
and there is no dynamical reason why they decay into open strings.
For $N$ large enough to suppress \NS-brane, but still
small enough to support strong interactions between closed strings, closed strings may decay into open strings.
We believe that further studies are needed to 
understand this phenomenon more systematically.

It is of note that 
the structure of the proposed phase diagram is 
reminiscent of known examples  
where systems flow into novel universality classes 
at interacting critical points.
For example, in the two-dimensional $Z_N$ clock model 
with $N$ greater than a critical value,
the critical point 
between the disordered phase 
and the ordered phase
has an emergent $U(1)$ symmetry\cite{ZN}.
More recently, it has been proposed that the critical
point between an antiferromagnetic state and a
valence bond state in 2+1 dimensions
can possess a non-trivial quantum order 
which supports an emergent gauge boson
and fractionalized excitations\cite{SENTHIL04}.

\subsection{World sheet description of deconfined string}

In order to make a contact with the traditional first quantization formulation of string theory, 
it will be useful to have a world-sheet description 
of deconfined strings in the holographic phases.
Here we focus on closed string.
The generalization to open string is straightforward.
Note that the hopping integral from loop $C_1$ 
to $C_1 + C_2$
is determined
by the loop field $\bar \phi_{C_2} = |\bar \phi_{C_2}| e^{i b_{C_2}}$ 
which is complex.
The amplitude $| \bar \phi_{C_2} |$ determines the strength of the hopping,
and defines the notion of `distance' between the two loops.
The distance between two loops, in turn, defines
the metric of the space in which loops are defined.
In this sense, condensates of loop fields determine 
the metric of the space
in which closed loops propagate.
The $U(1)$ phase $b_{C_2}$ corresponds to the
background two-form field 
to which closed strings are electrically coupled.



\begin{figure}[h!]
\centering
      \includegraphics[height=5cm,width=6cm]{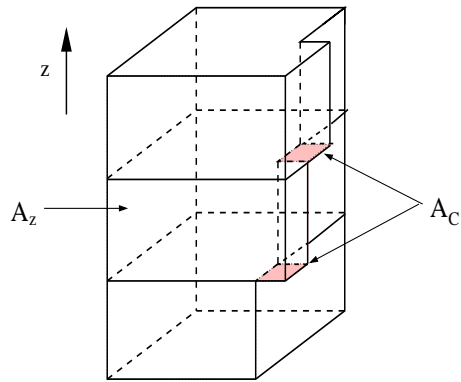}
\caption{
The world sheet action has two contributions :
$A_z$ is the area of the surface which is parallel to $z$,
and $A_C$ is the contribution from hoppings mediated by
condensates of loop fields.
}
\label{fig:loop_propagation2}
\end{figure}

This can be made more intuitive if we use a world-sheet representation.
Let us consider the the quadratic Hamiltonian that includes the tension and the hopping terms,
\bqa
H_0(z) & = & 
  L_C \chi_C^\dagger \chi_C 
 - \frac{ 1 }{M^2} \Bigl( 
 2 \bar \phi_{C_2}^{*}(z)  \chi_{C_1}^\dagger \chi_{C_1+C_2}
  + 2\bar  \phi_{C_2}(z)  \chi_{C_1+C_2}^\dagger \chi_{C_1} \Bigr).
\label{H0}
\eqa
Here we suppressed the form factors $F_{ij}(C_1,C_2)$ 
and $G_{ij}(C_1,C_2)$.
For this discussion, we ignore a possible
deformation of the path integral of the string fields.
We consider the single loop propagator given by
\bqa
g(C_2,z_2; C_1,z_1) & = & < C_2 | e^{- \int_{z_1}^{z_2} H_0(z) dz } | C_1 >,
\eqa
where $|C> = \chi_C^\dagger |0>$ is a single string state.
We can `re-discretize' the imaginary time $z$ 
into small steps with size $\epsilon$ to write 
\bqa
g(C_2,z_2; C_1,z_1) & = & \sum_{\mbox{all world sheets}} e^{-S_{WS}^1 + i S_{WS}^2 },
\eqa
where 
\bqa
S_{WS}^1  =  {\cal A}_z 
- \sum_{i} \ln ( 2  \epsilon |\bar \phi_{C_i}(z_i)| / M^2 ),
\eqa 
where ${\cal A}_z$ is the area of the world sheet whose faces
are tangential to the $z$ direction, and
the second term in $S_{WS}^1$ includes contributions 
$\frac{ 2 \epsilon |\bar \phi_{C_i}(z_i)|}{M^2}$ for each hopping
mediated by loop field $\phi_{C_i}$ at time $z_i$.
The second term is associated with the parts of the world sheet
that are perpendicular to the $z$ direction.
This is illustrated in Fig. \ref{fig:loop_propagation2}.
One can view $S_{SW}^1$ as the Nambu-Goto action
provided that the area associated with a loop $C$ at time $z$
is taken to be $A_C = -\ln ( 2 \epsilon |\bar \phi_C(z_i)| / M^2 )$.
The areas associated with loops in turn would determine 
the spatial metric of the space.
The imaginary part of the action,
\bqa
S_{WS}^2 = \int_{\mbox{world sheet}} B
\eqa
is simply the Berry phase associated 
with the phase of the background loop fields
in the temporal gauge.
Since both the phase and amplitude modes of loop fields
are dynamical, not only the two-form gauge field but also
a metric field should arise as a dynamical degree of freedom.

\begin{figure}[h!]
\centering
      \includegraphics[height=8cm,width=10cm]{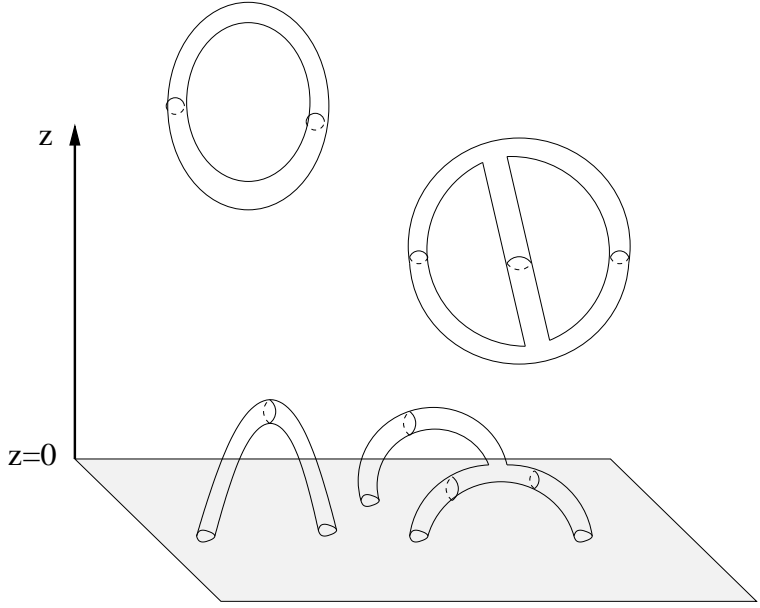}
\caption{
A snap shot of a vacuum fluctuation
in the holographic critical phase with deconfined closed string.
}
\label{fig:fluctuation}
\end{figure}

A typical configuration of vacuum fluctuation 
in the full interacting string theory are shown in 
Fig. \ref{fig:fluctuation}.
For every vertex where closed loops join or split,
there is a factor of $1/N$.
For a large $N$, 
the theory describes weakly interacting strings propagating
in the time-dependent background.

\subsection{Continuum limit}

In the holographic phase, there are three important length scales.
The first scale is associated with the tension $\mu_{NS_{D-4}}$ of \NS-brane, 
$l_{NS_{D-4}} \sim 1/(\mu_{NS_{D-4}})^{D-3}$.
The second scale is the `string scale' $l_s$ 
which corresponds to the typical size 
of closed string excitation in the bulk.
The third scale $L$ is the scale over which loop fields change appreciably in the bulk.
Roughly, the last one determines the `curvature' of the bulk space 
in which strings propagate.
One can take the continuum limit by tuning the system 
such that all these length scales are fixed in the $a \rightarrow 0$ limit.
If these scales satisfy $L >> l_s >> l_{NS_{D-4}}$, 
strings propagate in a weakly curved background.
It is expected that a continuum description of string theory 
emerges in this limit.
Since graviton has the same mass as the two-form gauge field in the continuum limit,
graviton may emerge as a massless mode 
along with the two-form gauge field 
in the holographic phase.
It will be interesting to see 
how the resulting theory in the continuum limit 
compares with the existing formulation of 
the closed string field theory\cite{ZWIEBACH}.


%
%

\section{Acknowledgment}

I would like to thank
G. Baskaran,
Umut Gursoy,
Sean Hartnoll,
Ellias Kiritsis,
Hong Liu,
John McGreevy,
Rob Myers
and
Subir Sachdev
for helpful comments and discussions.
This research was supported in part by 
the Natural Sciences and Engineering Research Council of Canada
and the Early Research Award from the Ontario Ministry of Research and Innovation.
Research at the Perimeter Institute is supported 
in part by the Government of Canada 
through Industry Canada, 
and by the Province of Ontario through the
Ministry of Research and Information.

\section{Appendix A}

We prove the identity
\bqa
I & = & \int d \phi^* d \phi ~~ e^{ - \phi ( \phi^* - A^* )} f(\phi^*) \nn
&=& \pi f(A^*)
\eqa
for any analytic function $f(x)$.
We define a new variable $\xi = \phi - A$ to write
\bqa
I & = & 
\int d \xi^* d \xi ~~ e^{ - |\xi|^2 - A \xi^* } f( \xi^* + A^* ) \nn
& = & 
\sum_{m=0}^\infty \sum_{n=0}^\infty \frac{1}{m! n!} 
\int d \xi^* d \xi ~~ e^{ - |\xi|^2 } ( - A \xi^* )^m  f^{(n)}(A^*) \xi^{*n},
\eqa
where we Taylor expanded $e^{A \xi^*}$ and $f(\xi^*+A^*)$.
Here only $m=n=0$ component survives in the angular integration of $\xi = |\xi| e^{i \theta}$
and we have
\bqa
I & = & \pi f(A^*).
\eqa

\section{Appendix B}

Here we show that Eq. (\ref{transition})
with Eqs. (\ref{Psi_i}) - (\ref{wf_f}) is 
equivalent to Eq. (\ref{Z_main0}).
The imaginary time $\beta$ is divided into $R$ pieces,
\bqa
Z & = & \int \prod_{l=1}^{R-1} d\Phi_C^{(l)*} d\Phi_C^{(l)} ~
< \Psi_f| e^{-\epsilon H} |\Phi_C^{(R-1)}> e^{-\Phi_C^{(R-1)*} \Phi_C^{(R-1)} } 
<\Phi_C^{(R-1)}| e^{-\epsilon H} |\Phi_C^{(R-2)}> \times \nn
&& e^{-\Phi_C^{(R-2)*} \Phi_C^{(R-2)} }  ...
<\Phi_C^{(2)}| e^{-\epsilon H} |\Phi_C^{(1)}> e^{-\Phi_C^{(1)*} \Phi_C^{(1)} } 
<\Phi_C^{(1)}| e^{-\epsilon H} |\Psi_i>, 
\eqa
where we use the identity
\bqa
\int d\Phi_C^{(l)*} d\Phi_C^{(l)} ~~
|\Phi_C^{(l)}> e^{-\Phi_C^{(l)*} \Phi_C^{(l)} } <\Phi_C^{(l)}| = 1.
\eqa
For (\ref{Psi_i}) and (\ref{Psi_f}), the transition amplitude becomes
\bqa
Z & = & \int \prod_{l=0}^{R} d\Phi_C^{(l)*} d\Phi_C^{(l)} ~~
\Psi_f^*[\Phi_C^{(R)*},\Phi_C^{(R)}]
\Psi_i[\Phi_C^{(0)*},\Phi_C^{(0)}] ~~
e^{-S},
\eqa
where
\bqa
S & = & - \Phi_C^{(R)*} \Phi_C^{(R)} 
+ \sum_{l=1}^R \left[ \Phi_C^{(l)*} ( \Phi_C^{(l)} - \Phi_C^{(l-1)} ) + \epsilon H( \Phi_C^{(l)*}, \Phi_C^{(l-1)} )  \right].
\label{Sa}
\eqa
We take $R \rightarrow \infty$ limit
and equate Eq. (\ref{Sa}) with Eq. (\ref{Z_main0}) to identify
\bqa
\Psi_i[\Phi_C^{*},\Phi_C^{}] & = & e^{ - S_{UV}[\Phi_C^*/N,\Phi_C/N] }, \nn
\Psi_f^*[\Phi_C^{*},\Phi_C^{}] & = & e^{ -\Phi_C^* \Phi_C
-  S_{IR}[\Phi_C/N] }.
\eqa

\section{Appendix C}

\begin{figure}[h!]
\centering
      \includegraphics[height=4cm,width=4cm]{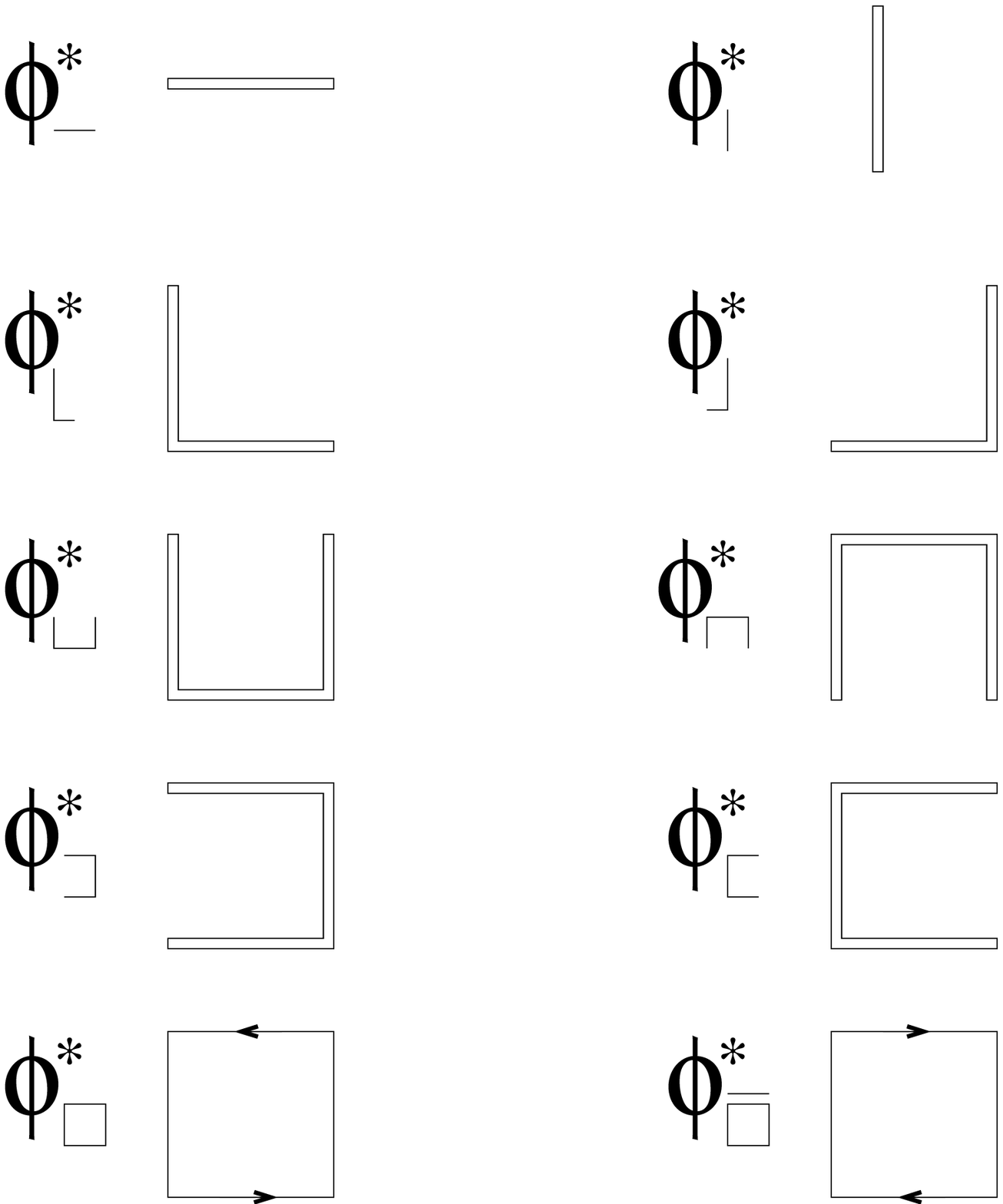}
\caption{
}
\label{fig:symbols}
\end{figure}

In this section, we solve the loop equations perturbative in $1/M$ to the lowest non-trivial order for small loops.
Let us first consider the saddle point solution in the large $M$ limit,
which corresponds to the deep inside the confinement phase
of the gauge theory.
For $M=\infty$, we have simple solution,
\bqa
\phi_C(z)  =  \phi_C(0) e^{-L_C z}, ~~~
\phi_C^*(z)  =  \phi_C^*(0) e^{L_C z}
\eqa
with the boundary conditions in Eqs. (\ref{BC_UV}) and (\ref{BC_IR}).
Especially, the IR boundary condition implies
$\phi_C^*(\infty) = 0$, which then implies
$\phi_C^*(z)=0$.
This makes sense physically 
because the source for Wilson loops decreases exponentially with $z$,
and the expectation values of Wilson loop operators vanish
in the strong coupling limit of the gauge theory.
For a large but finite $M$, the loop equations 
can be solved perturbatively in $1/M$. 

To illustrate the idea, 
we solve the loop equation for a simple model
which include only the single trace operator
for the unit plaquette, $V=J_\Box \phi_\Box^*$.
In this model, we solve $\phi_\Box^*$ to the lowest order in $1/M$.
Keeping the first four shortest loops that contribute 
to the evolution of $\phi_C^*$ of unit plaquette,
we write the saddle point equations,
\bqa
\partial_z \phi_\Box^* & = & 4 \phi_\Box^* 
- \frac{2}{M^2} ( \phi_\sqcup^* + \phi_\sqcap^* + \phi_\sqsubset^* + \phi_\sqsupset^* ) \phi_{\bar \Box}, \nn 
\partial_z \phi_\sqcup^* & = & 6 \phi_\sqcup^* 
- \frac{2}{M^2} ( \phi_\lfloor^* + \phi_\mid^{*2} + \phi_\rfloor^* ), \nn 
\partial_z \phi_\lfloor^* & = & 4 \phi_\lfloor^* 
- \frac{2}{M^2} ( \phi_\mid^* + \phi_-^* ), \nn 
\partial_z \phi_-^* & = & 2 \phi_\Box^* 
- \frac{2}{M^2}. 
\eqa
The symbols are defined in Fig. \ref{fig:symbols}.
We consider solutions with the translational and discrete rotational symmetries
of the lattice, and suppress the site index for the loop fields.
The equations for $\phi_\sqcap^*$, $\phi_\sqsubset^*$ and $\phi_\sqsupset$
are similar to the one for $\phi_\sqcup^*$.
Similarly, the equation for $\phi_\rfloor^*$ ($\phi_\mid$)
is similar to the one for $\phi_\lfloor^*$ ($\phi_-$).
From Eqs. (\ref{Vp}) and (\ref{BC_IR}), 
we read the boundary conditions,
\bqa
\phi_\Box^*(\infty)  =   \frac{\phi_{\bar \Box}(\infty)}{M^8}, ~~~  
\phi_\sqcup^*(\infty)   =   \frac{1}{M^6}, \nn 
\phi_\lfloor^*(\infty)   =   \frac{1}{M^4}, ~~~
\phi_-^*(\infty)   =  \frac{1}{M^2}. 
\eqa
Using the zero-th order solution for $\phi_{\bar \Box} = J_{\bar \Box} e^{-4z}$,
we obtain
\bqa
\phi_{\Box}^*(z) = \frac{J_{\bar \Box}}{M^8} e^{-4z}.
\eqa
This implies that the expectation value of the Wilson loop operator for the unit plaquette
become $J_{\bar \Box}/M^8$ as expected in the large $M$ limit.
The difference from the case with $M=\infty$ is that 
$\phi_C^*$ is now non-zero at $z=0$ 
because Wilson loops have non-zero expectation values.
As expected, $\phi_C$ and $\phi_C^*$ 
satisfy Eq. (\ref{BC_IR2}) to the leading order in $1/M$ at all $z$.

\end{document}